\documentclass{aa}  

\usepackage{graphicx}
\usepackage{txfonts}
\usepackage{xcolor}
\usepackage{ulem}
\usepackage{pgffor}
\usepackage{xcolor}
\usepackage{rotating}
\usepackage{tikz}
\usepackage{morefloats}
\usepackage{morefloats}
\usepackage{graphicx}
\usepackage{caption}
\usepackage{subcaption}
\usepackage{csvsimple}
\usepackage{longtable}
\usepackage{booktabs}
\usepackage{pdfpages}
\usepackage{pgffor}
\usepackage{multicol}
\usepackage{footnote}
\usepackage{etoolbox}
\usepackage{array} 
\usepackage{multirow}
\usepackage{threeparttable}
\usepackage[pdftex]{hyperref}

\begin{document} 

\newcommand{\znfe}{[Zn/Fe]$_{\mathrm{fit}}$}
\newcommand{\znfei}{[Zn/Fe]$_{\mathrm{fit}, i}$}

   \title{Connecting the dusty dots: dust depletion and extinction of local interstellar clouds}

   \author{T. Ramburuth-Hurt \inst{1,2}
   \and
           A. De Cia \inst{3}
    \and
            J.-K. Krogager \inst{4,5}
    \and
            C. Ledoux \inst{6}
    \and
        A. J. Fox \inst{7} }

    \institute{Department of Astronomy, University of Geneva, Chemin Pegasi 51, Versoix, Switzerland 
    \and
    Wits Centre for Astrophysics, School of Physics, University of the Witwatersrand, 1 Jan Smuts Avenue, Johannesburg, 2000, South Africa 
    \and
        European Southern Observatory, Karl-Schwarzschild-Str. 2, 85748 Garching, Germany 
    \and
        Université Claude Bernard Lyon 1, Centre de Recherche Astrophysique de Lyon UMR5574, 9 Av. Charles André, 69230 Saint-Genis-Laval, France 
    \and
        French-Chilean Laboratory for Astronomy (FCLA), CNRS-IRL3386, U. de Chile, Camino el Observatorio 1515, Casilla 36-D, Santiago, Chile 
    \and 
        European Southern Observatory, Alonso de Córdova 3107, Vitacura, Casilla 19001, Santiago, Chile 
    \and 
        AURA for ESA, Space Telescope Science Institute, 3700 San Martin Drive, Baltimore, MD 21218, USA 
    }
   \date{Received xx; accepted yy}

  \abstract  
{Investigating the chemical complexity of the interstellar medium (ISM) is key
for understanding its physical nature and evolution. In this work, we study
parsec-scale interstellar dust clouds in the neutral ISM of the Milky Way using
two different probes: dust depletion and dust extinction. We examine their
relationship to investigate the distribution of metals and dust in the Solar
neighbourhood, and how they are related to the Local Bubble. We use measurements
of dust depletion for individual gas clouds along eight lines of sight towards
bright O/B stars within 1.1 kpc of the Sun, derived from UV absorption-line
spectra. We combine these with parsec-scale 3D dust extinction density maps out
to 1.25 kpc. Based on the well-known relationship between gas and dust in the
ISM, we assume a correlation between dust depletion and dust extinction density,
which we use to imply that the absorption components are spatially associated
with the peaks in dust extinction density, and to pinpoint the likely locations
of the gas clouds in physical space. Using the Python \texttt{scipy} package
\texttt{\texttt{find\_peaks}}, we identify peaks in the dust extinction curves,
and then associate the stronger peaks with the strongest dust depletion
components. Independent distance measurements along the line of sight towards
one of our targets, $\theta^1$ Ori C, validates our result and supports the
reliability of our method. In our sample, the minimum distance between clouds
that have significantly different chemical properties (in terms of dust
depletion) is $\sim$ 100pc. This gives an indication on the physical scale on
which chemical mixing remains incomplete in the ISM of the Milky Way. For five
of the eight targets, we report dust depletion values for gas clouds associated
with the Local Bubble. Additionally, we find a velocity gradient that is
consistent with the expansion of the Local Bubble, further supporting our
methodology. Overall, we show that it is possible to use complementary
information from dust depletion and dust extinction to build more detailed maps
of ISM metal and dust distributions.}


   \maketitle

\section{Introduction}

Studies of the chemical composition of the Milky Way's interstellar medium (ISM)
gives us important insight into its evolution and the role of different gaseous
phases in this environment. Both dust depletion and dust extinction are probes
of the dust content in the ISM of galaxies. Dust extinction is a measure of how
much light is absorbed by the dust in all gas phases of the ISM along the line
of sight, including the cold and dense gas. In the Milky Way, dust extinction
can be determined for different positions in the sky, and, with the inclusion of
photometry and astrometry with Gaia, it is possible to construct
three-dimensional dust maps \citep{Lallement+2018, Green+2019, Rezaei+2024,
Dharmawardena+2024, Edenhofer+2024}. 

Absorption-line spectroscopy is a powerful method for studying the metal content
of the gas in galaxies because it gives access to the column densities of many
metals (ions), and the methods of measuring the column densities are highly
robust. Additionally, UV absorption-line spectroscopy is particularly valuable
because the resonant lines of the most dominant ionisation states of the most
abundant cosmic metals (C, O, Mg, Si, S, Fe, etc) are available in the UV. 

Dust depletion is the phenomenon whereby metals are incorporated into dust
grains and are no longer observable in the gas phase \citep{Field1974,
Savage&Sembach1996, Jenkins2009}. The depletions of different metals correlate
with each other to varying degrees depending on how easily they form dust. The
measure of the relative abundances between metals with different refractory
properties is therefore a measurement of the amount of dust depletion. In a
method for characterising dust depletion, called the \textit{relative method}
\citep{DeCia+2016, DeCia+2021, Konstantopoulou+2022} the determination of the
refractory indices of metals is based on relative metal abundances. While the
initial implementation of the relative method have used [Zn/Fe] as the main
tracer of the overall strength of dust depletion, the most recent
implementations use all relative abundances that trace dust depletion
\citep{Konstantopoulou+2024}. An advantage of this method is that it does not
make any assumption on the metallicity of the gas. An additional advantage is
that with high-resolution absorption-line spectroscopy it is possible to dissect
lines of sight and identify and study the chemical properties of individual gas
components \citep{Welty+2020, Ramburuth-Hurt+2023, Ramburuth-Hurt+2025}.
Differences in dust depletion in individual gas components along lines of sight
are washed away by the more typical studies of dust depletion, which study the
integrated line of sight \citep[e.g.][]{Jenkins2009, DeCia+2021,
Roman-Duval+2021, Ritchey+2023}.

The relationship between the dust probed by dust depletion and that probed
through dust extinction is not well-understood. Several works that compare dust
extinction based on dust depletion measured from full lines of sight
($A_{V,~depl}$) and extinction ($A_{V,~ext}$) in extra-galactic damped
Ly-$\alpha$ absorbers \citep[DLAs, e.g.][]{Savaglio&Fall2004, Wiseman+2017,
Bolmer+2019} often find significant scatter. Recently,
\cite{Konstantopoulou+2024} developed a more robust methodology for calculating
$A_{V,~depl}$ which uses all metals and the relative method for the integration
of the dust-to-metal ratio (DTM). They find that, for integrated line of sight
measurements, $A_{V,~depl}$ and $A_{V,~ext}$ generally follow each other (see
figure 7 of that paper), with the exception of systems that have the 2175\AA\,
bump in their extinction curve, possibly associated with polycyclic aromatic
hydrocarbon dust particles (PAHs). In these cases, \cite{Konstantopoulou+2024}
find that $A_{V,~ext} > A_{V,~depl}$. This inequality could be caused by a large
amount of PAHs in the cold neutral medium (CNM), which is captured by dust
extinction measurements but not by dust depletion. Dust depletion traces the
dust that can be probed by metal absorption lines present in the warm neutral
medium (WNM). In the Milky Way, the 2175 \AA\, bump is generally ubiquitous
\citep{Savage&Mathis1979,Pei1992,Gordon+2003}. 

Three-dimensional dust maps of the Milky Way make it possible to locate regions
of the ISM with high dust content along lines of sight \citep{Lallement+2018,
Green+2019, Leike+2020, Rezaei+2024, Dharmawardena+2024, Edenhofer+2024}. These
maps have enabled detailed studies of phenomena such as the Local Bubble,
including its geometry and origin. The Local Bubble is a supernova-driven
low-density cavity, a few hundred parsecs in diameter, in which the Sun resides
\citep{Fuchs+2006, Zucker+2022, ONeill+2024}. \cite{Zucker+2022} show that the
several nearby star-forming regions are located on the surface of the Local
Bubble. A recent study by \cite{ONeill+2024}, based on the 3D dust map of
\cite{Edenhofer+2024}, shows a chimney-like structure, likely produced by the
bursting of the supernova-driven bubble. 

Direct comparisons between dust extinction and dust depletion in individual ISM
components along lines of sight have, to our knowledge, not been done before.
This is likely because both the method of calculating the dust depletion for
individual components \citep{DeCia+2016, Ramburuth-Hurt+2023,
Ramburuth-Hurt+2025} and the availability of high spatial resolution 3D dust
extinction maps \citep[e.g.][]{Lallement+2018, Green+2019, Leike+2020,
Rezaei+2024, Dharmawardena+2024, Edenhofer+2024} are relatively recent.

In this paper, we compare these two tracers of dust -- dust depletion and dust
extinction -- between ISM components to investigate their relationship and the
possibility of cross-matching strong features in the dust extinction density
with components identified from metal absorption lines. We use the dust
depletion measurements of individual gas components along eight lines of sight
from \citet[][henceforth R-H+25]{Ramburuth-Hurt+2025} and the 3D dust extinction
density maps from \citet[][henceforth E+24]{Edenhofer+2024}. 
    
\section{Data} \label{sec:data}

\subsection{Dust depletion}

The dust depletion measurements for individual (groups of) components
[Zn/Fe]$_{i, \mathrm{fit}}$ are taken from the eight lines of sight in R-H+25
(see Table \ref{tbl:metal-patterns-results-all-targets}). Here, gas components
refer to contributions from gas clouds along the line of sight to the overall
metal absorption profile, each of which can be fitted with a Voigt profile in
velocity space. Because the decomposition of lines is not unique, R-H+25 merged
individual components into groups with similar velocities such that their total
column densities could be robustly determined. The choice of grouping can be
somewhat arbitrary and is a conservative representation of the distribution of
individual components along the line of sight. We describe this in more detail
in Section \ref{sec:methods}.

\subsection{Dust extinction density}

We obtain dust extinction density distributions from E+24, who map the dust
extinction density within 1.25 kpc of the Sun. These maps have an angular
resolution of 14 arcsec and parsec-scale distance resolution. From the library
of available 3D dust maps, we choose E+24 due to its small angular resolution,
making it most comparable to the pencil-beam measurements from absorption-line
spectroscopy. E+24 is also based on the \cite{Zhang+2023} catalogue from the
most recent Gaia data release \citep[DR3][]{GaiaCollaboration2023} out to 1.25
kpc, which contains all R-H+25 targets. This catalogue has distance and
extinction measurements for 220 million stars, with the advantage of having
small uncertainties ($< 0.06$ mag) on the extinction for 87 million of these.

\cite{Zhang+2023} report the parameter ``extinction'' in terms of $E$, which has
units of magnitudes and is roughly equal to the reddening $E(B-V)$. Because
\cite{Zhang+2023} assume a universal extinction curve $R(\lambda)$
\footnote{Available at \url{https://zenodo.org/records/7811871}}, the value of
$E$ can be translated to an extinction value at a given wavelength $\lambda$.
This means that $E$ is therefore essentially a scalar reddening factor that is
applied to the stellar spectrum through the assumed $R(\lambda)$. We access the
E+24 dust extinction density curves, i.e. $E$ per parsec as a function of
heliocentric distance, using the \texttt{dustmaps} Python module
\citep{Green2018}, in which we query for each line of sight by their Galactic
coordinates.

\begin{table*}
\centering
\caption{Details of the eight lines of sight studied in this paper, including the dust depletion measurements for individual (groups of) components from R-H+25. }
\begin{threeparttable}
\begin{tabular}{c|c|c|c|c|c|c}
\hline \hline
\textbf{Target}  & \textbf{Distance (pc)}$^{[1]}$ & \textbf{Galactic \textit{l, b} (deg)}   & \textbf{$A_V$ (mag)}$^{[2]}$ & \textbf{Group} & \textbf{Dust depletion}    & \textbf{Velocity range (km s$^{-1}$)} \\ 
\hline
\multirow{4}{*}{$\theta^1$ Ori C} & \multirow{4}{*}{$393^{+18}_{-20}$} & \multirow{4}{*}{209.01, $-$19.38} & \multirow{4}{*}{1.78$\pm$0.36} &  1 & 1.30$\pm$0.15 & $[-2.1,14.1]$\\
                                   &                                   &                                  & & 2 & 1.46$\pm$0.20 & $[16.6,25.4]$\\
                                   &                                   &                                  & & 3 & 0.74$\pm$0.10 & $[28.0,32.0]$\\
                                   &                                   &                                  & & 4 & 0.60$\pm$0.21 & $[35.5,40.2]$\\ \hline
\multirow{2}{*}{HD~110432}        & \multirow{2}{*}{$439\pm15$} & \multirow{2}{*}{302.0, $-$0.20}      & \multirow{2}{*}{--}             & 1 & 1.80$\pm$0.59 & $[-10.1,-6.4]$\\
                                   &                                   &                                  & & 2 & 1.84$\pm$0.06 & $[-1.0,8.2]$\\ \hline
\multirow{2}{*}{$\rho$ Oph A$^{\dagger}$} & \multirow{2}{*}{$137\pm3$} & \multirow{2}{*}{353.69, 17.69} & \multirow{2}{*}{2.58$\pm$0.34} & 1 & 1.32$\pm$0.18 & $[-29.2,-20.4]$\\
                                   &                                   &                                  & & 2 & [2.0]          & $[-13.4,-0.3]$\\ \hline
\multirow{4}{*}{$\chi$ Oph}       & \multirow{4}{*}{$152\pm5$} & \multirow{4}{*}{357.93, 20.68}       & \multirow{4}{*}{--}             & 1 & 1.03$\pm$0.13 & $[-33.0,-22.4]$\\
                                   &                                   &                                  & & 2 & 1.79$\pm$0.04 & $[-20.8,-15.3]$\\
                                   &                                   &                                  & & 3 & 2.03$\pm$0.13 & $[-12.7,-6.4]$\\
                                   &                                   &                                  & & 4 & 1.70$\pm$0.16 & $[-4.5,0.7]$\\ \hline
\multirow{5}{*}{HD~154368}        & \multirow{5}{*}{$1030^{+38}_{-33}$} & \multirow{5}{*}{349.97, 3.22} & \multirow{5}{*}{2.53$\pm$0.20} & 1 & 1.58$\pm$0.08 & $[-27.7,-20.7]$\\
                                   &                                   &                                  & & 2 & 1.62$\pm$0.20 & 15\\
                                   &                                   &                                  & & 3 & 0.80$\pm$0.20 & $[-11.3,-9.3]$\\
                                   &                                   &                                  & & 4 & 1.85$\pm$0.09 & $[-6.5,0.1]$\\
                                   &                                   &                                  & & 5 & 1.22$\pm$0.27 & $[4.2,13.1]$\\ \hline
\multirow{5}{*}{$\kappa$ Aql} & \multirow{5}{*}{$506^{+53}_{-41}$} & \multirow{5}{*}{31.77, $-$13.29} & \multirow{5}{*}{--} & 1 & 0.96$\pm$0.11 & $[-28.2,-18.2]$\\
                               &                                   &                     & & 2 & 1.74$\pm$0.09 & $[-16.2,-11.1]$\\
                               &                                   &                     & & 3 & 1.60$\pm$0.06 & $[-8.7,-6.2]$\\
                               &                                   &                     & & 4 & 1.09$\pm$0.28 & $[-2.1,1.5]$\\
                               &                                   &                     & & 5 & 0.63$\pm$0.06 & $[5.8,13.9]$\\ \hline
\multirow{2}{*}{HD~206267} & \multirow{2}{*}{$790_{-113}^{+171}$} & \multirow{2}{*}{99.29, 3.74} & \multirow{2}{*}{1.47$\pm$0.14} & 1 & 1.41$\pm$0.07 & $[-32.1,-36.7]$\\
                           &                                   &                                & & 2 & 1.2$\pm$0.05  & $[-14.6,-5.3]$\\ \hline
\multirow{3}{*}{HD~207198} & \multirow{3}{*}{$978_{-27}^{+34}$} & \multirow{3}{*}{103.14, 6.99} & \multirow{3}{*}{1.50$\pm$0.29} & 1 & 1.25$\pm$0.25 & $[-34.3,-27.8]$\\
                           &                                   &                                & & 2 & 1.55$\pm$0.14 & $[-22.1,-9.8]$\\
                           &                                   &                                & & 3 & 0.87$\pm$0.49 & 4.7\\ \hline

\end{tabular}
\begin{tablenotes}
\footnotesize
\item[$^{\dagger}$] The square brackets for the second component along the line of sight towards $\rho$ Oph A indicate that the dust depletion measurement for this component is unconstrained due to the saturation of most metal absorption lines.
\item [1] Distances are from the Sun and taken from Gaia DR3 \citep{Bailer-Jones+2021}
\item [2] \citet{Valencic+2004}
\end{tablenotes}
\end{threeparttable}
\label{tbl:metal-patterns-results-all-targets}
\end{table*}

\section{Methods} \label{sec:methods}

Our aim is to investigate whether we can pinpoint the locations of the (groups
of) absorbing gas clouds identified in absorption by cross-matching these to
peaks in dust extinction density maps. We adopt the column densities and dust
depletion measurements for individual components \znfei\, as they are presented
in R-H+25. 

We smooth the dust extinction density maps using a Gaussian kernel of 4 pc in
order to reduce the impact of noise in the peak detection process, and we set
the detection prominence at 2.04 $\times~10^{-6}~E$/pc \citep[following the
parameters in][]{ONeill+2024}. We then identify the peaks using the
\texttt{scipy} python package \texttt{find\_peaks}. \cite{ONeill+2024} use this
methodology to construct the geometry of the Local Bubble, where they define the
shell of the Local Bubble as the first significant peak (with Gaussian kernel of
7 pc and prominence of 2.04 $\times~10^{-6}~E$/pc) in extinction density along
lines of sight. We find the same first peaks as \cite{ONeill+2024} using a
Gaussian kernel of 4 pc, with the benefit of detecting relevant additional peaks
beyond the first.

Along the line of sight to $\theta^1$ Ori C, we include an additional ``peak''
at the stellar distance. Although there is no distinct peak in the dust
extinction density profile, there is a sharp increase in this region, indicating
the existence of a dense dust cloud. We consider the dust extinction density at
the upper end of the stellar distance uncertainty as a lower limit to ensure
that we capture the dust contribution accurately.

We attempt to convert the UV components from velocity space to heliocentric
distance under the assumption that their kinematics are due to a flat Galactic
rotation curve $R = 234.88 / \omega_R$ \citep{Clemens1985}, with $R_{\odot} =
8.15$ kpc and $v_{\odot} = 236$ km/s \citep{Reid+2019}. We find no plausible
physical solutions from this exercise, likely due to the clouds being very close
by, therefore making the their relative motions a result of local effects and
not global Galactic rotation. All our targets are also on the same side of the
Milky Way, and thus do not rotate relative to each other. 

To quantify the relationship between dust depletion of absorbing gas components
and peaks in dust extinction density, we assume that dust and gas are broadly
correlated, so that the absorption components with the highest level of dust
depletion trace the strongest dust extinction peaks. This assumption is based on
the well-known relationship between gas and dust in the ISM \citep{Bohlin+1978,
Cardelli+1989}, which gives the average value of the Galactic gas-to-dust ratio,
$N(\mathrm{H}$\textsc{i}$)/E(B-V)= 5.8 \times 10^{21}$ cm$^{-2}$ mag$^{-1}$. We
expect the correlation between dust depletion [Zn/Fe]$_{\mathrm{fit, }i}$ and
dust extinction density to be stronger than that between [Zn/Fe]$_{\mathrm{fit,
}i}$ and dust extinction $A_V$ itself. This is because depletion represents an
average number of dust particles per unit length of absorbing cloud along the
line of sight, making it more related to the density of dust extinction than to
dust extinction, which is the integrated amount of dust.

Then, for each line of sight we arrange both quantities in descending order and
pair them accordingly, so that the strongest dust depletion component is paired
with the strongest dust extinction component, and so on. A consistent comparison
is ensured by including only the top $n$ values of each, where $n$ is the
smaller number of components or peaks along the line of sight. We exclude the
line of sight towards $\rho$ Oph A because it has only one constrained dust
depletion component due to saturation of many of its metal absorption lines. We
then infer the locations of the absorption components in physical space. 

We estimate the minimum distance over which chemical mixing in the ISM is
incomplete by comparing pairs of absorbing gas components along each line of
sight and determining whether their dust depletion \znfei\, differ
significantly. We determine the statistical significance of the difference in
\znfei\, for components $A$ and $B$ using a z-test: $\sigma_{\mathrm{z-test}} =
\frac{\mathrm{[ZnFe]}_{i,A} - \mathrm{[ZnFe]}_{i,B}}{\sqrt{Err(
\mathrm{[ZnFe]}_{i,A})^2 + Err(  \mathrm{[ZnFe]}_{i,B})^2}}$, where
$\mathrm{[ZnFe]}_{i,A}$ and $Err~(\mathrm{[ZnFe]}_{i,A})$ are the values for
dust depletion its uncertainty. We consider two components to have significantly
different \znfei\, if $\sigma_{\mathrm{z-test}} > 3$. 

\section{Results and discussion} \label{sec:results}

We find an overall agreement between the number of (groups of) absorption
components and the number of peaks in dust extinction density, i.e. within $\pm$
1 -- 2, for six of the eight lines of sight in our sample ($\theta^1$ Ori C,
HD~110432, $\rho$ Oph A, HD~154368, $\chi$ Oph and $\kappa$ Aql, see Table
\ref{tab:znfe_av_components}). For the lines of sight towards both HD~206267 and
HD~207198, there are many more extinction peaks than absorption components.
There are several possible explanations for this. The broader field of view in
dust extinction measurements compared to the narrow pencil-beam nature of
absorption-line spectroscopy can cause extinction to include more clouds along
the line of sight. It is also likely that peaks in dust extinction density may
correspond to regions at the same velocity, hence they become blended in the
absorption profiles. Extinction also typically captures the contribution from
the dust present in the CNM, e.g. PAHs. Our method of determining dust depletion
is not sensitive to PAHs and carbonaceous dust because the carbon absorption
line is generally saturated. Another explanation is that components with lower
levels of dust depletion are lost in the noise, or that they have a less
significant contribution to the overall dust compared to PAHs. We also note that
our grouping of absorption components is conservative and therefore does not
fully represent the substructure of the gas along lines of sight. 

\begin{figure}
    \centering
    \includegraphics[width=\linewidth]{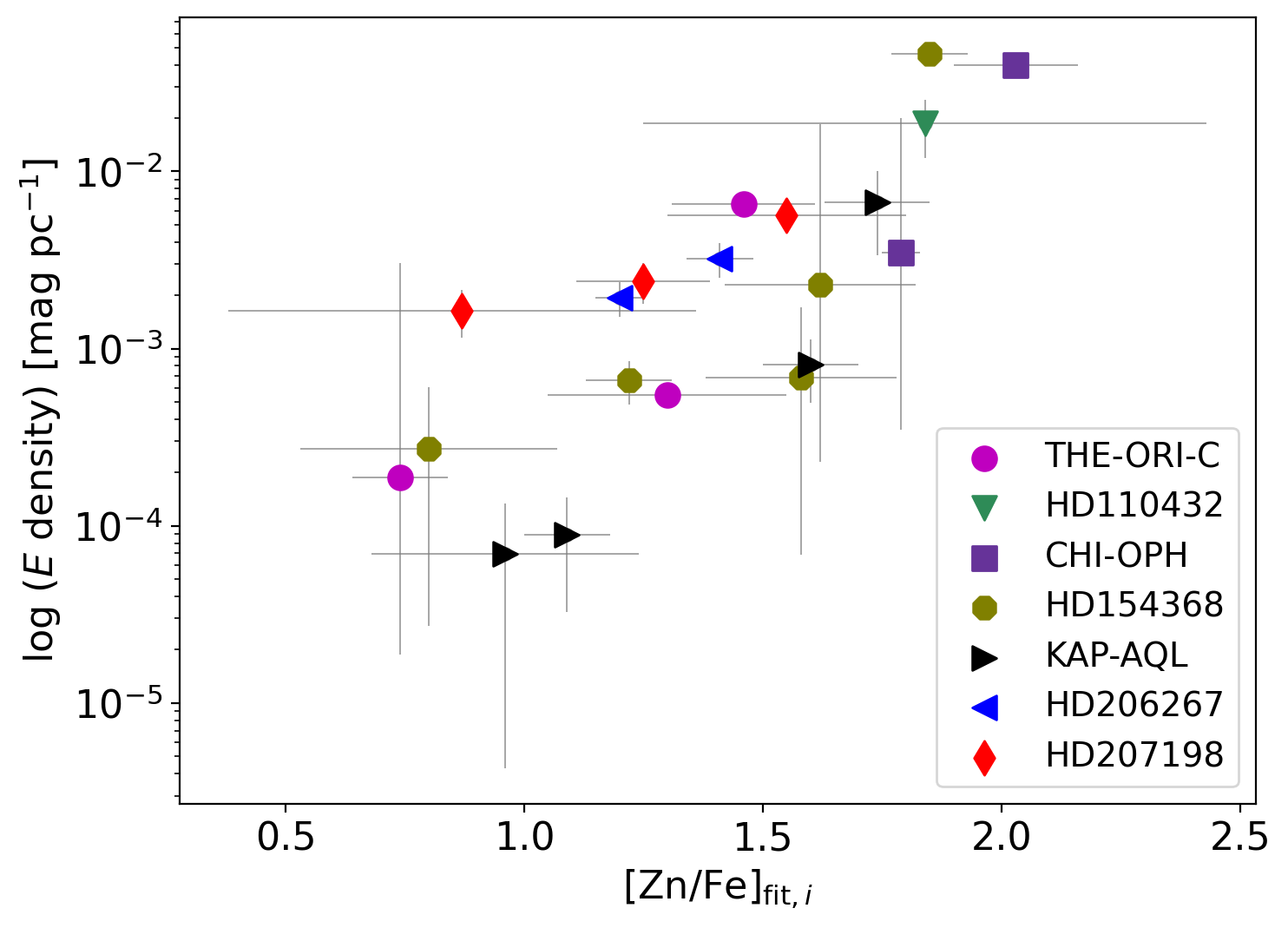}
    \caption{Comparison of dust extinction density with level of dust depletion \znfei~ for individual (groups of) gas components along all lines of sight in our sample.}
    \label{fig:depl-vs-ext}
\end{figure}

We find a relatively small scatter in the overall relation between dust
depletion and dust extinction density across all lines of sight, shown in Fig.
\ref{fig:depl-vs-ext}. Although we somewhat build in the correlation between
dust depletion and dust extinction density along each line of sight by assigning
the components with the strongest depletion to the clouds with the highest dust
extinction density, the normalisation and slope of these correlations are free
to vary and may, in principle, be different from line of sight to line of sight.
However, we find a strong Spearman rank coefficient of 0.81 (p $= 1.6 \times
10^{-5}$) across all lines of sight, which indicates that components with higher
dust depletion likely correspond to higher dust extinction densities. The fact
that all lines of sight individually display similar positive correlations is
encouraging and likely represents a physical link between the two parameters. We
also find a moderate Pearson correlation coefficient of 0.63 (p $=$ 0.003)
between dust depletion and dust extinction density, which suggests that the
relationship between dust depletion and dust extinction density is unlikely to
be linear. 

To quantify the significance of the scatter between \znfei\, and log
($E$ density), as shown in Figure \ref{fig:depl-vs-ext}, we perform a
Kolmogrov-Smirnov (KS) test on the normalised orthogonal residuals of the
best-fit line to the dust extinction density vs dust depletion. We find a
p-value from the KS test of $p_{KS} = 0.79$, which indicates that the normalised
orthogonal residuals are consistent with being drawn from a normal Gaussian
distribution. Therefore, the scatter can be fully explained by the noise in the
data. We provide the full analysis in Appendix \ref{appendix:scatter_analysis}.

We use the correlation between dust extinction density and dust depletion to
then infer that the locations of the gas cloud coincide with that of the dust
extinction density peak along each line of sight in descending order (see
Section \ref{sec:methods}), which we present in Table
\ref{tab:znfe_av_components}. Some distances are larger than the corresponding
stellar distances Table \ref{tbl:metal-patterns-results-all-targets}, but these
reflect the combined uncertainties of the stellar distances and dust extinction
maps.

For the line of sight towards $\theta^1$ Ori C, we find that the region with the
highest dust extinction density is around 411 pc, which coincides with the
distance of the Orion Nebula itself. Furthermore, the absorption component with
the highest level of dust depletion (here Component 2 with velocity range 17 --
25 km s$^{-1}$ and [Zn/Fe]$_{\mathrm{fit},2} = 1.46 \pm 0.20$), also resides
within the Nebula. This has been shown through independent studies of absorption
lines towards both $\theta^1$ Ori C, which is a star within the Orion Nebula,
and the nearby star $\iota$ Ori, which is in the foreground of the Nebula
\citep[][Dalla Pola 2024 Master Thesis]{Price+2001}. The agreement between the
distance to the highest extinction density peak and the absorption component
with the highest level of dust depletion strongly supports our methodology.
R-H+2025 show that this absorption component likely also has a high gas fraction
and super-Solar metallicity. 

Components 1 and 3 along the line of sight towards $\theta^1$ Ori C also have
the smallest distance between each other with a statistically significant
($\sigma_{\mathrm{z-test}} = 3.1$) difference in dust depletion levels. With
Component 1 at 159 pc with [Zn/Fe]$_{\mathrm{fit},2} = 1.46 \pm 0.20$ and
Component 3 at 257 pc with [Zn/Fe]$_{\mathrm{fit},3} = 0.74 \pm 0.10$, their
difference in dust depletion is $0.56 \pm 0.25$ over a distance of 98 pc. We
interpret this as the distance where the chemical mixing is not complete. While
it could be even smaller, we show that this is an effective method to give
meaningful constraints on the physical scale of gas, metal and dust mixing.
Indeed, \citep{Redfield&Linksy2008} find significant differences in the
gas-phase abundances of Fe and Mg between the interstellar gas clouds, Local
Interstellar Cloud and Cloud G, which are separated by only 1 pc. They attribute
these differences to dust depletion, and could also be an indication of the
distance at which chemical mixing is not complete.

\begin{table}
\centering
\begin{threeparttable}
\caption{\znfei\, values with their corresponding dust extinction densities and distance from the Sun for each line of sight.}
\label{tab:znfe_av_components}
\begin{tabular}{llll} 
\hline \hline
\textbf{Target} & \textbf{\znfei} & \textbf{log $E$ density} & \textbf{Distance} \\
                &                 & (mag pc$^{-1}$)         & (pc)               \\
\hline
\multirow{4}{*}{$\theta^1$ Ori C} 
  & 1.46 $\pm$ 0.20 & $> -2.18$        & $\sim$ 411.0  \\
  & 1.30 $\pm$ 0.15 & $-3.26 \pm 0.23$ & 159.0 $\textit{(LB)}$  \\
  & 0.74 $\pm$ 0.10 & $-3.73 \pm 0.18$ & 257.0 \\
  & 0.60 $\pm$ 0.21 & --                 & --     \\
\hline
\multirow{2}{*}{HD~110432}
  & 1.84 $\pm$ 0.06 & $- 1.73 \pm 0.16 $ & 196.5 $\textit{(LB)}$ \\
  & 1.80 $\pm$ 0.59 & --                 & --  \\
\hline
\multirow{2}{*}{$\rho$ Oph A $^{\dagger}$}
  & [2.0]            & $- 2.44 \pm 0.26 $ & 121.2 $\textit{(LB)}$ \\ 
  & 1.32 $\pm$ 0.18  & --                 & --     \\ 
\hline
\multirow{4}{*}{$\chi$ Oph}
  & 2.03 $\pm$ 0.13  & $- 1.4 \pm 0.18 $  & 150.1  \\
  & 1.79 $\pm$ 0.04  & $- 2.46 \pm 0.23 $ & 113.9 $\textit{(LB)}$ \\
  & 1.70 $\pm$ 0.16  & --                 & --   \\
  & 1.03 $\pm$ 0.13  & --                 & --   \\
\hline
\multirow{6}{*}{HD~154368}
  & 1.85 $\pm$ 0.09 & $- 1.33 \pm 0.15 $ & 201.2  \\
  & 1.62 $\pm$ 0.20 & $- 2.64 \pm 0.19 $ & 259.0  \\
  & 1.58 $\pm$ 0.08 & $- 3.16 \pm 0.22 $ & 154.1 $\textit{(LB)}$ \\
  & 1.22 $\pm$ 0.27 & $- 3.18 \pm 0.22 $ & 898.9  \\
  & 0.80 $\pm$ 0.20 & $- 3.56 \pm 0.29 $ & 509.4  \\
  & --              & $- 4.0 \pm 0.29 $ & 1061.6  \\
\hline
\multirow{5}{*}{$\kappa$ Aql}
  & 1.74 $\pm$ 0.09 & $- 2.17 \pm 0.22 $ & 135.6 $\textit{(LB)}$ \\
  & 1.60 $\pm$ 0.06 & $- 3.09 \pm 0.17 $ & 281.2  \\
  & 1.09 $\pm$ 0.28 & $- 4.05 \pm 0.28 $ & 466.0  \\
  & 0.96 $\pm$ 0.11 & $- 4.16 \pm 0.41 $ & 527.6  \\
  & 0.63 $\pm$ 0.06 & --                    & --  \\
\hline
\multirow{6}{*}{HD~206267 $^{\dag\dag}$}
  & 1.41 $\pm$ 0.04 & $- 2.49 \pm 0.2 $ & 461.5  \\
  & 1.20 $\pm$ 0.05 & $- 2.71 \pm 0.16 $ & 247.9 $\textit{(LB)}$ \\
  & --              & $- 2.96 \pm 0.17 $ & 340.2 \\
  & --              & $- 3.14 \pm 0.3 $ & 515.7 \\
  & --              & $- 3.32 \pm 0.2 $ & 606.2 \\
  & --              & $- 3.48 \pm 0.18 $ & 722.0  \\
\hline
\multirow{7}{*}{HD~207198 $^{\ddag}$}
  & 1.55 $\pm$ 0.14 & $- 2.25 \pm 0.18 $ & 425.0 \\
  & 1.25 $\pm$ 0.25 & $- 2.62 \pm 0.32 $ & 616.3 \\
  & 0.87 $\pm$ 0.49 & $- 2.78 \pm 0.16 $ & 227.6 \\
  & --              & $- 3.12 \pm 0.28 $ & 368.6 \\
  & --              & $- 3.34 \pm 0.29 $ & 541.8 \\
  & --              & $- 3.73 \pm 0.26 $ & 177.2 $\textit{(LB)}$ \\
  & --              & $- 3.93 \pm 0.24 $ & 912.4 \\
\hline
\end{tabular}
\tablefoot{
    \tablefoottext{LB}{The component associated with the Local Bubble.}
    \tablefoottext{$\dagger$}{We tabulate the results for $\rho$ Oph A here but note that we have not associated either absorption component with dust extinction density peaks for this line of sight.}
    \tablefoottext{$\dag\dag$}{We also tabulate the results for HD~206267 and HD~207198 but the significant mismatch between the number of absorption components and dust extinction density peaks makes them difficult to interpret.}
}
\end{threeparttable}
\end{table}

\begin{figure}[]
    \centering
    \includegraphics[width=0.9\linewidth]{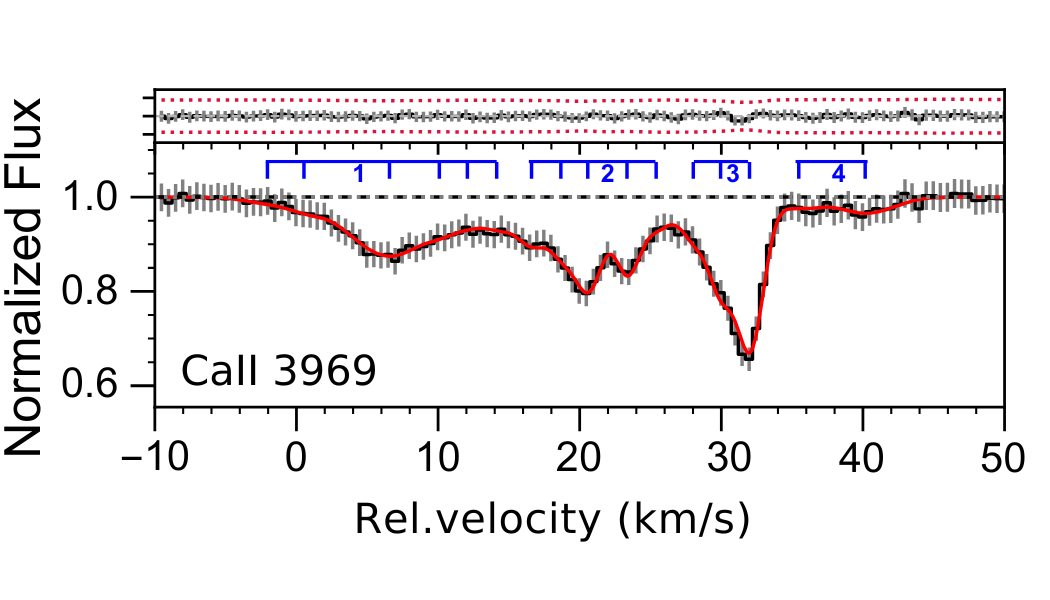}

    \includegraphics[width=0.9\linewidth]{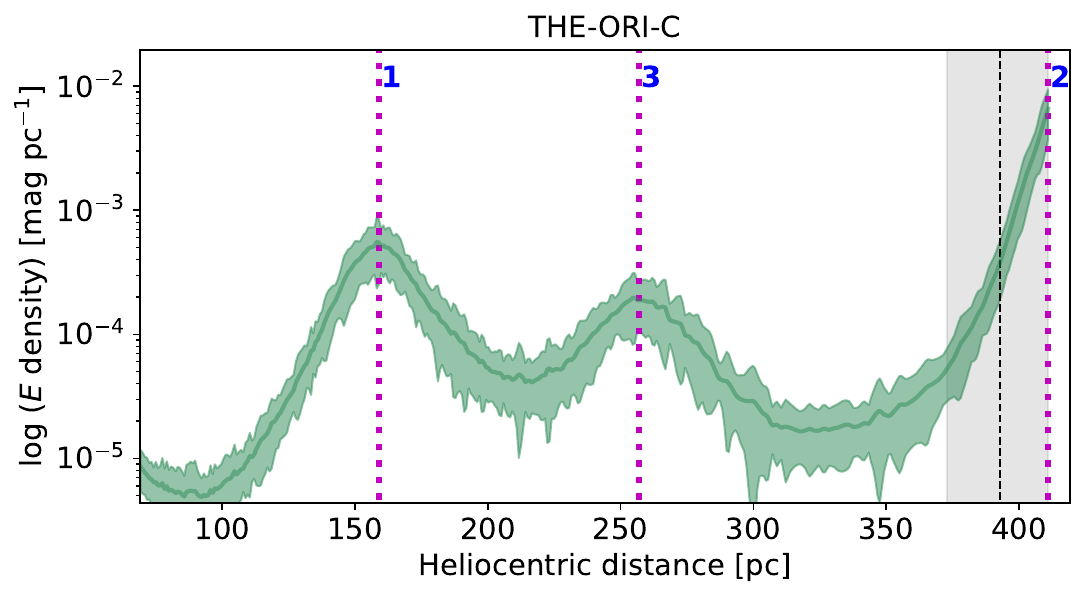}\\ [-1.6mm] 

    \caption{Top: Ca\,\textsc{ii} absorption line for $\theta^1$ Ori C are shown by the solid black line in the lower panel of this figure. The solid red lines are the fitted Voigt profiles, which were obtained with \texttt{VoigtFit} \citep{Krogager2018_VoigtFit} in R-H+25. The blue ticks represent the velocity-space positions of each individual component, which have then been grouped, indicated by the blue vertical lines. The values for \znfei\, given in Table \ref{tbl:metal-patterns-results-all-targets} correspond to the absorption groups in these figures as numbered from left to right in blue. The $1\sigma$ residuals are plotted as the black line in the upper panel of this figure, with $\pm ~3\sigma$ marked by the horizontal dotted lines. Bottom: Dust extinction density curve in log space for the same target. The vertical dotted pink lines show the locations of the extinction density peaks. The vertical dashed black line represents the location of the star, with the uncertainties as the shaded region. The blue numbers correspond to the respective absorption components, according to our method of matching absorption components to peaks in dust extinction density (see Section \ref{sec:methods}). }
    \label{fig:THE-ORI-C}
\end{figure}

\begin{figure}[]
    \centering
    \includegraphics[width=0.9\linewidth]{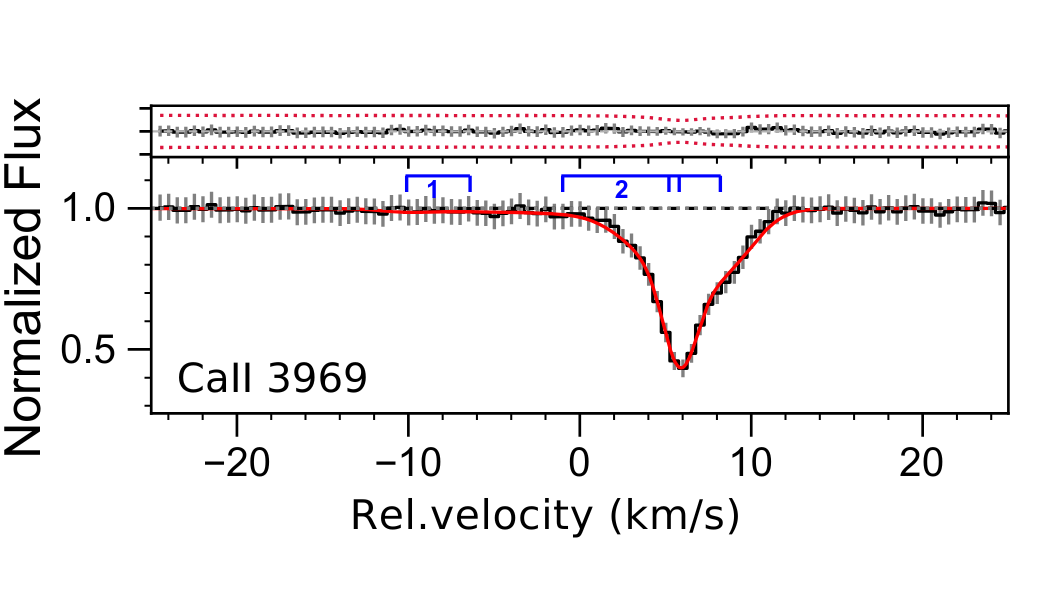}
    \includegraphics[width=0.9\linewidth]{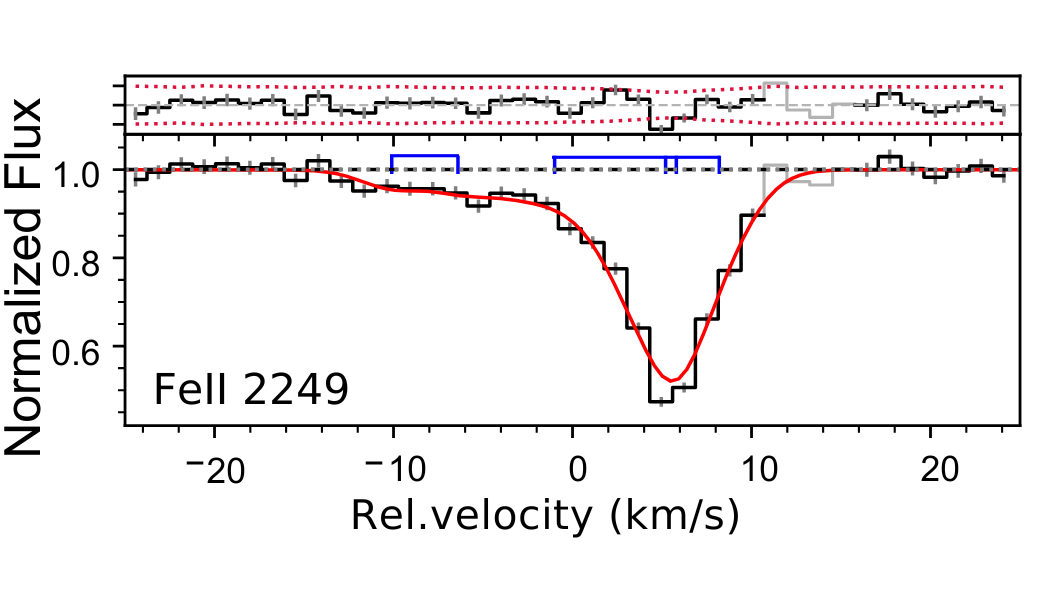}\\[-2mm]

    \includegraphics[width=0.9\linewidth]{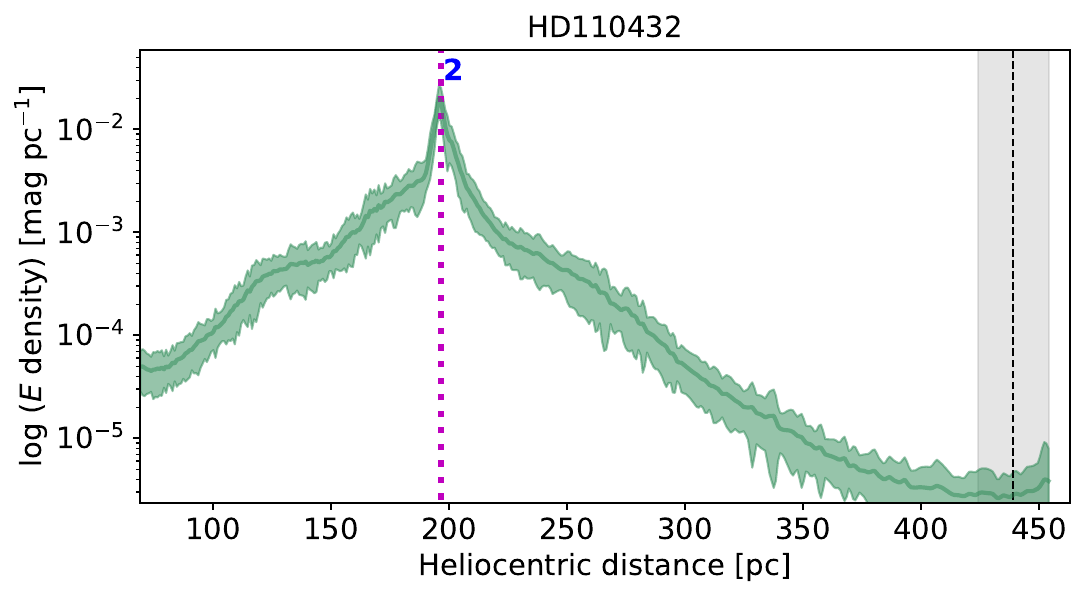}\\[-1.6mm] 

    \caption{Same as Fig. \ref{fig:CHI-OPH} and including the Fe\,\textsc{ii} $\lambda$2249 line for HD~110432. }
    \label{fig:HD110432}
\end{figure}

\begin{figure}[]
    \centering
    \includegraphics[width=0.9\linewidth]{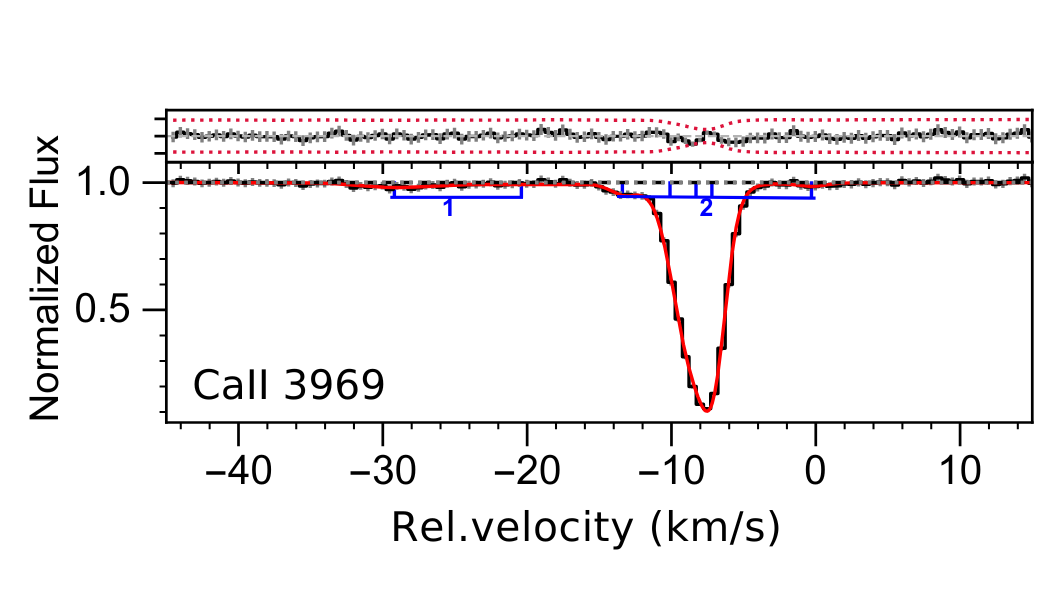}
    \includegraphics[width=0.9\linewidth]{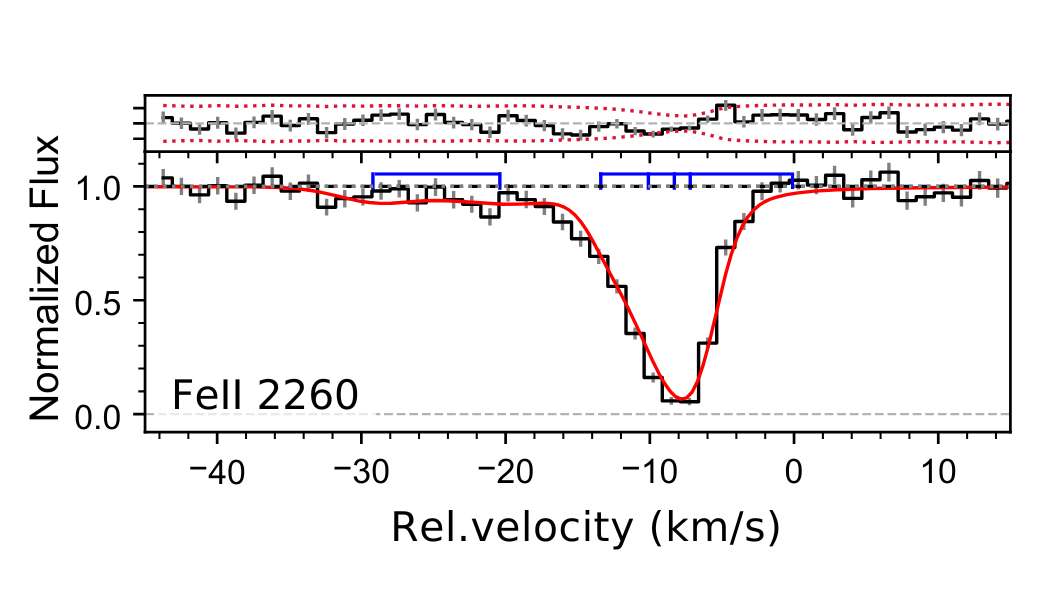}
    
    \includegraphics[width=0.9\linewidth]{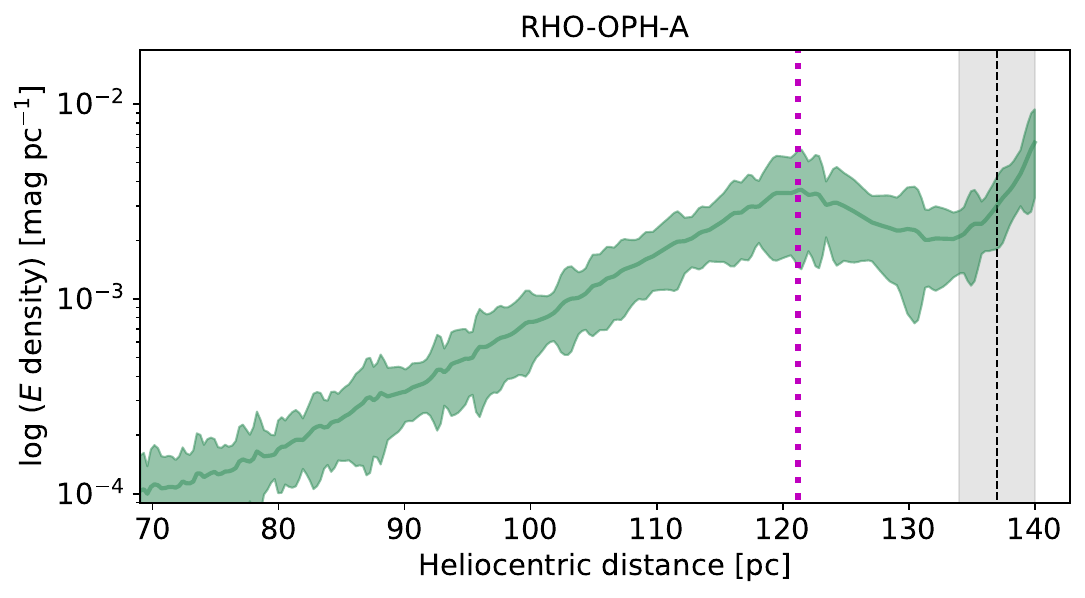}\\[-1.6mm]
    
    \caption{Same as Fig. \ref{fig:CHI-OPH} and including the Fe\,\textsc{ii} $\lambda$2260 line for $\rho$ Oph A. The dust extinction density curve has not been numbered in this case because we do not assign absorption components to dust extinction density peaks for this line of sight. }
    \label{fig:RHO-OPH-A}
\end{figure}

\begin{figure}[]
    \centering
    \includegraphics[width=0.9\linewidth]{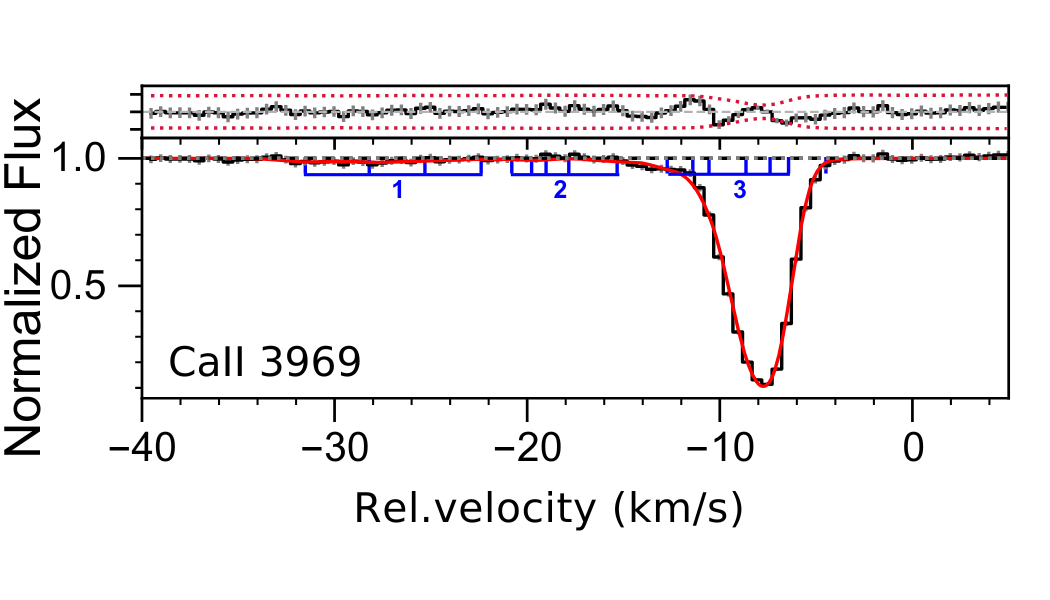}
    \includegraphics[width=0.9\linewidth]{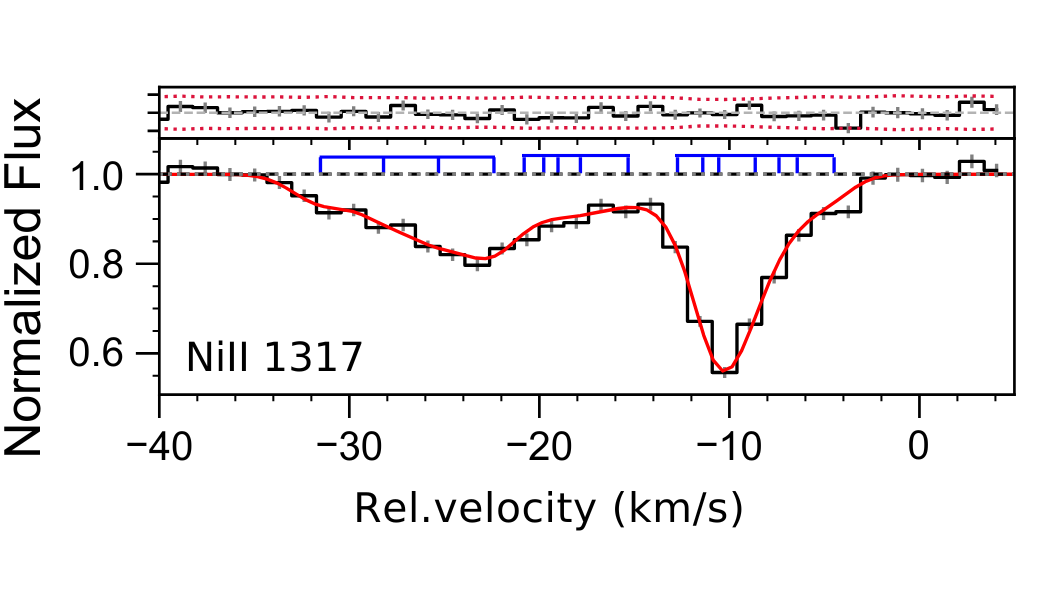}

    \includegraphics[width=0.9\linewidth]{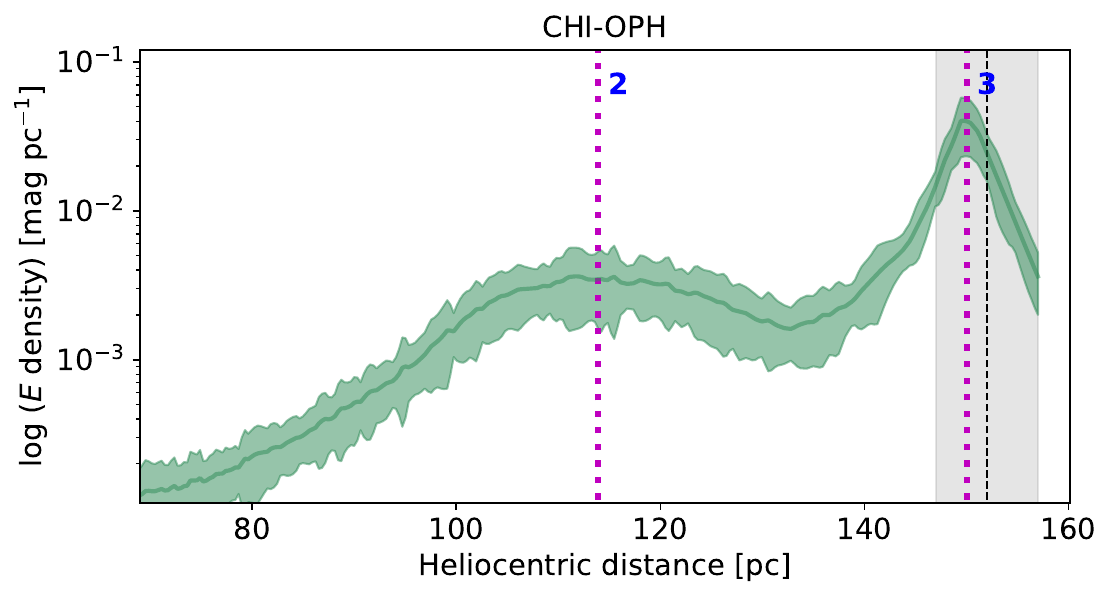}\\[-1.6mm] 

    \caption{Same as Fig. \ref{fig:THE-ORI-C} and including the Ni\,\textsc{ii} $\lambda$1317 line for $\chi$ Oph.}
    \label{fig:CHI-OPH}
\end{figure}

\begin{figure}[]
    \centering
    \includegraphics[width=0.9\linewidth]{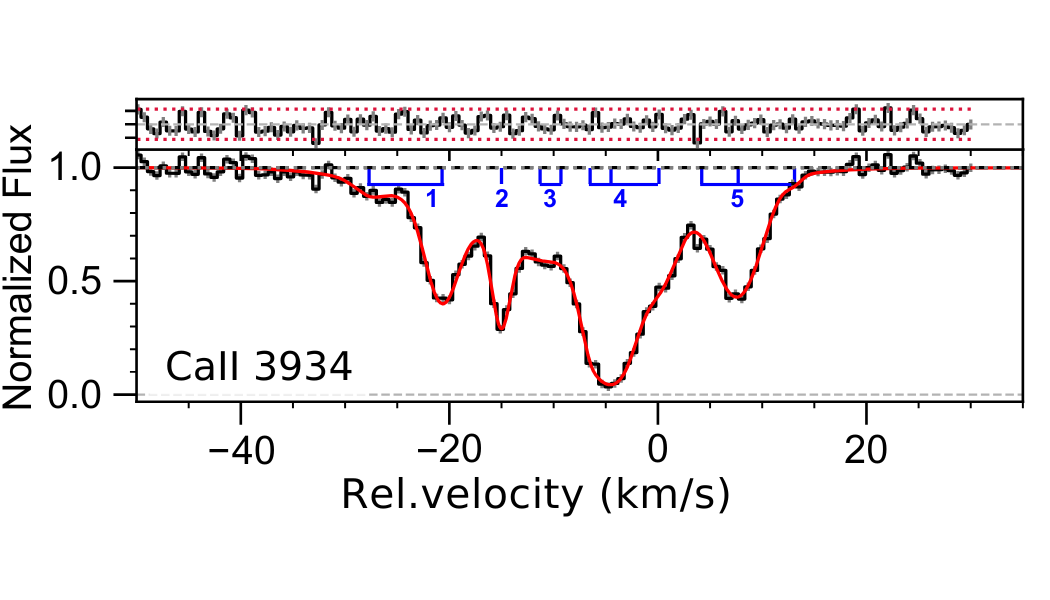}
    
    \includegraphics[width=0.9\linewidth]{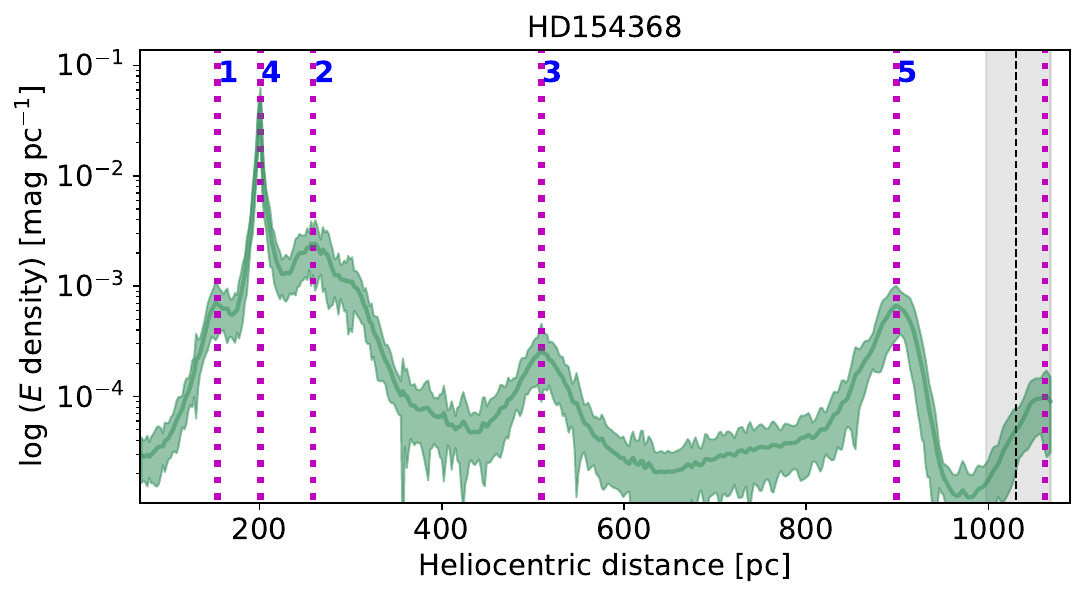}\\[-1.6mm]  
    
    \caption{Same as Fig. \ref{fig:THE-ORI-C} for HD~154368.}
    \label{fig:HD154368}
\end{figure}

\begin{figure}[]
    \centering
    \includegraphics[width=0.9\linewidth]{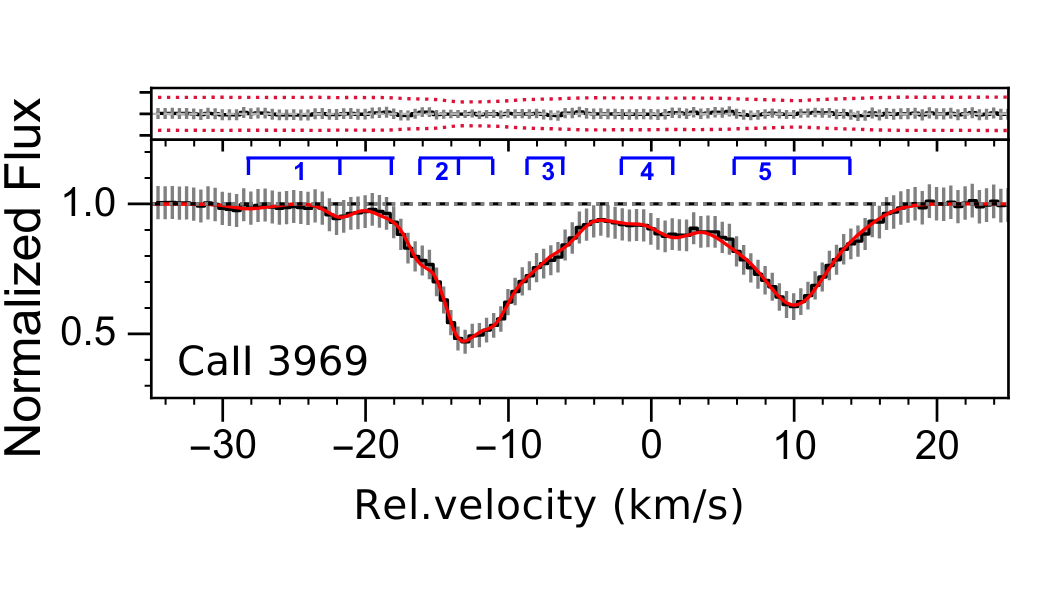}

    \includegraphics[width=0.9\linewidth]{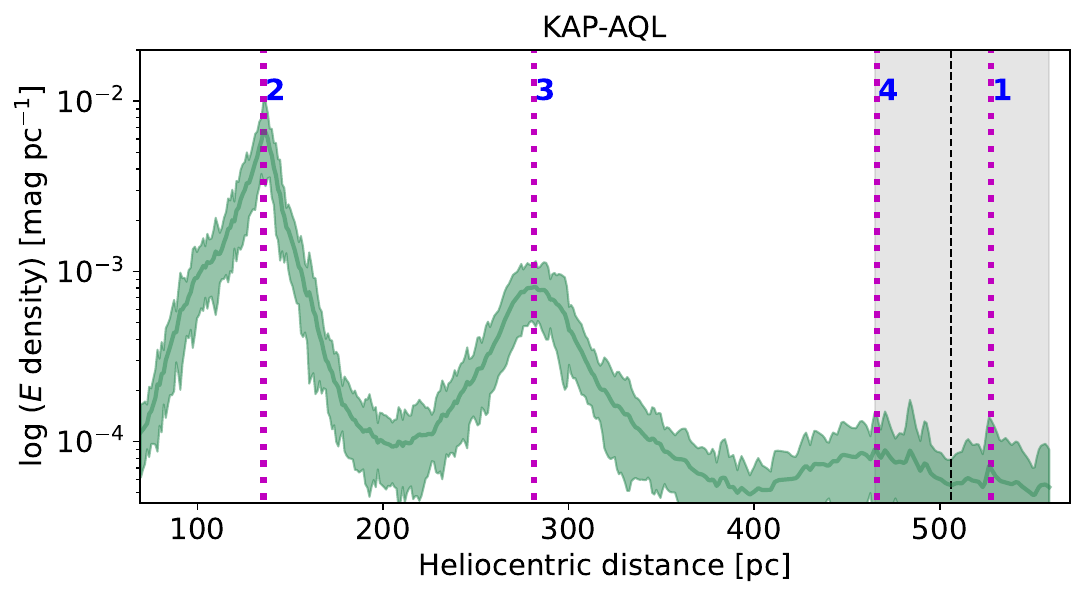}\\[-1.6mm] 

    \caption{Same as Fig. \ref{fig:THE-ORI-C} for $\kappa$ Aql.}
    \label{fig:KAP-AQL}
\end{figure}

\begin{figure}[]
    \centering
    \includegraphics[width=0.9\linewidth]{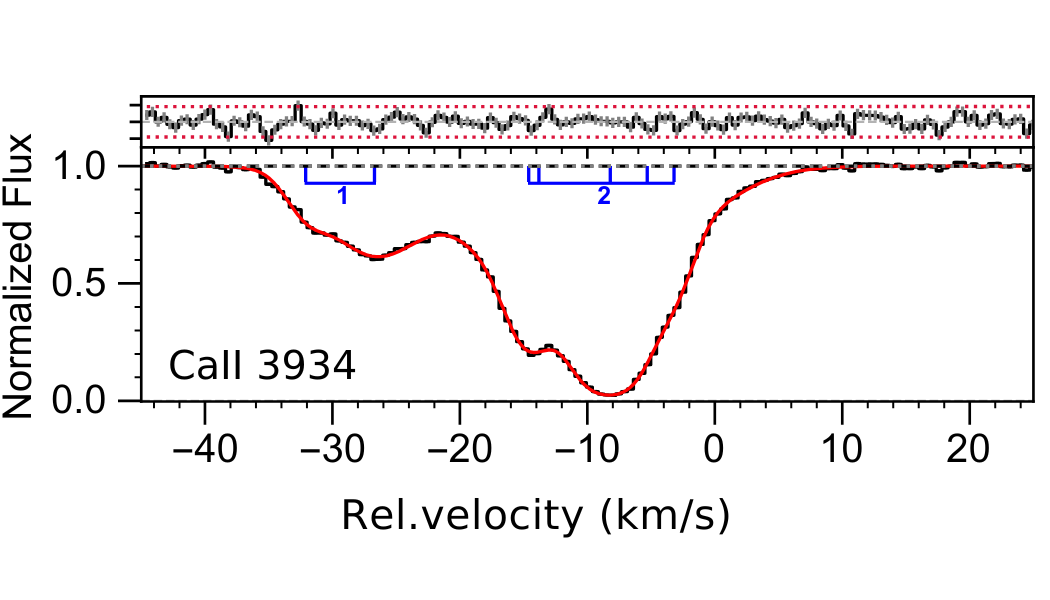}

    \includegraphics[width=0.9\linewidth]{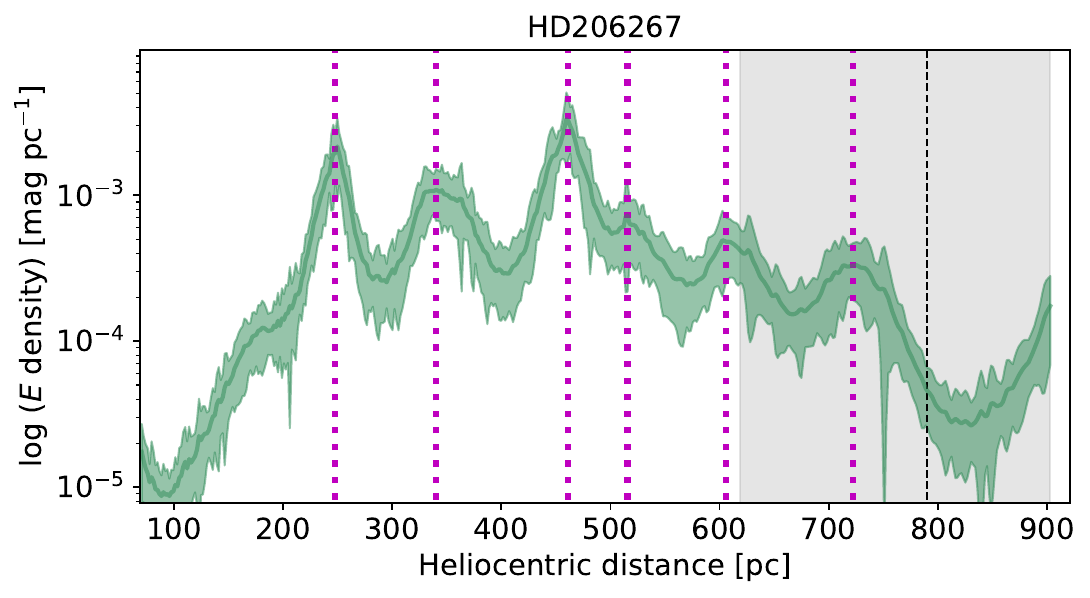}\\[-1.6mm]

    \caption{Same as Fig. \ref{fig:THE-ORI-C} for HD~206267. The dust extinction density curve has not been numbered in this case because the significant mismatch between the number of absorption components and peaks in dust extinction density makes interpretation unclear. }
    \label{fig:HD206267}
\end{figure}

\begin{figure}[]
    \centering
    \includegraphics[width=0.9\linewidth]{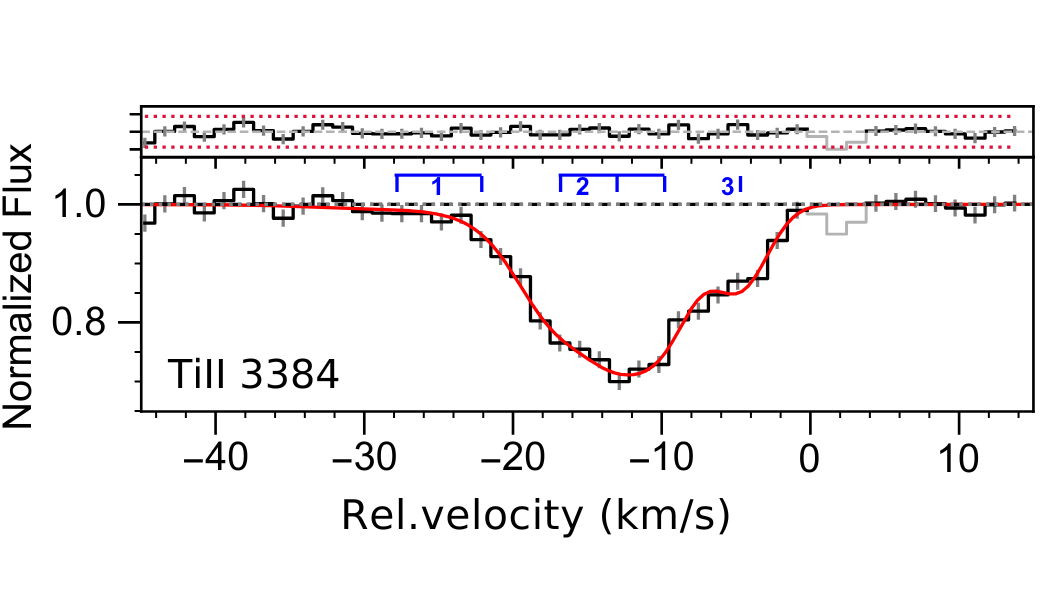}
    
    \includegraphics[width=0.9\linewidth]{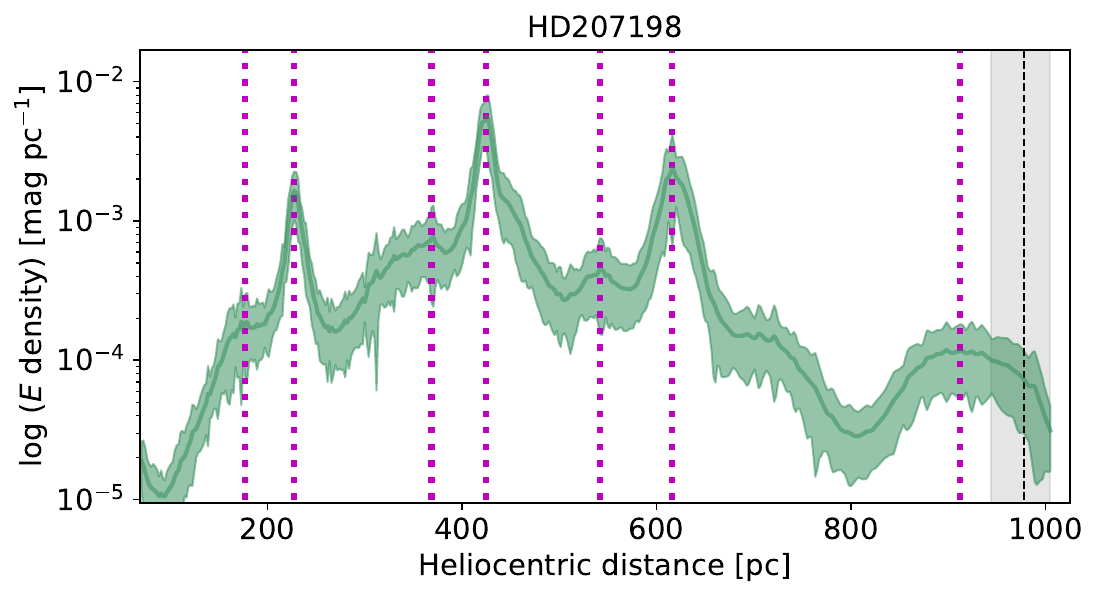}\\[-1.6mm] 
    
    \caption{Same as Fig. \ref{fig:THE-ORI-C} for HD~207189. Similarly to HD~206267, the dust extinction density curve has not been numbered due to the significant mismatch between the number of absorption components and peaks in dust extinction density making it difficult to interpret.}
    \label{fig:HD207189}
\end{figure}

In the top panels of Figs. \ref{fig:THE-ORI-C} -- \ref{fig:HD207189} we show the
absorption line profiles for Ca\,\textsc{ii} $\lambda$3934 or $\lambda$3969,
where available for each line of sight, taken from R-H+25. These spectra have
the highest spectral resolution and therefore most clearly demonstrate the
individual components. The spectra for $\theta^1$ Ori C, HD~110432, $\chi$ Oph,
$\rho$ Oph A and $\kappa$ Aql come from VLT/ESPRESSO (ID: 0102.C-0699(A), $R
\sim 190 000$, FWHM $\sim$ 1.54 km s$^{-1}$). For HD~154368 and HD~206267, the
spectra are from the Kitt Peak National Observatory (KPNO) Coude Feed Telescope
Camera 6 ($R \sim 220~000$, FWHM $\sim$ 1.35 km s$^{-1}$). For HD~207198, we use
the Ti\,\textsc{ii} $\lambda$3384 line due to the unavailability of
Ca\,\textsc{ii}, which comes from VLT/UVES (ID 194.C-0833, $R \sim 92 600$, FWHM
$\sim$ 3.2 km s$^{-1}$). For some targets, we also include an additional metal
absorption line to account for components that are very weak in the
Ca\,\textsc{ii} line. These come from HST/STIS (ID: 16750, $R \sim 114~000$,
FWHM $\sim$ 2.63 km s$^{-1}$). 

\begin{figure}[]
    \centering
    \includegraphics[width=0.9\linewidth]{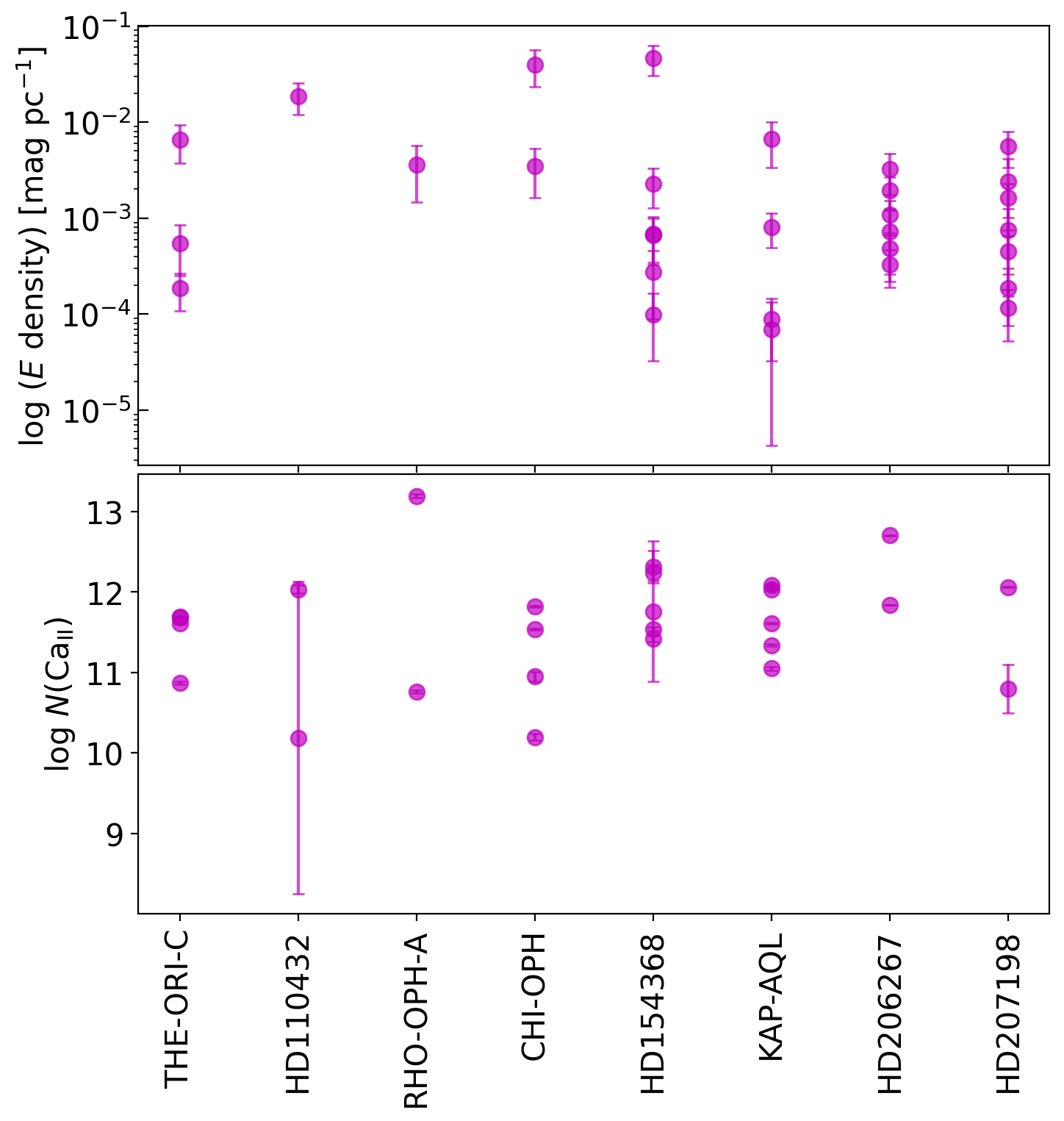}
    \caption{Top: Distribution of dust extinction densities along each line of sight. Bottom: Distributions of Ca\,\textsc{ii} column densities.}
    \label{fig:distributions}
\end{figure}

The bottom panels of Figs. \ref{fig:THE-ORI-C} -- \ref{fig:HD207189} show the dust extinction density curves along each line of sight from E+24 in log scale. The peaks in dust extinction density are plotted as vertical dashed lines (also see Table \ref{tab:znfe_av_components}). As mentioned in Section \ref{sec:methods}, it is possible that some broader peaks are the combination of two or more close-by to each other. We therefore take the number of peaks counted as lower limits. To show the overall distribution of dust extinction densities, we plot the dust extinction density of each identified peak along our lines of sight in Fig. \ref{fig:distributions}. 

\begin{figure*}
    \centering
    \includegraphics[width=\linewidth]{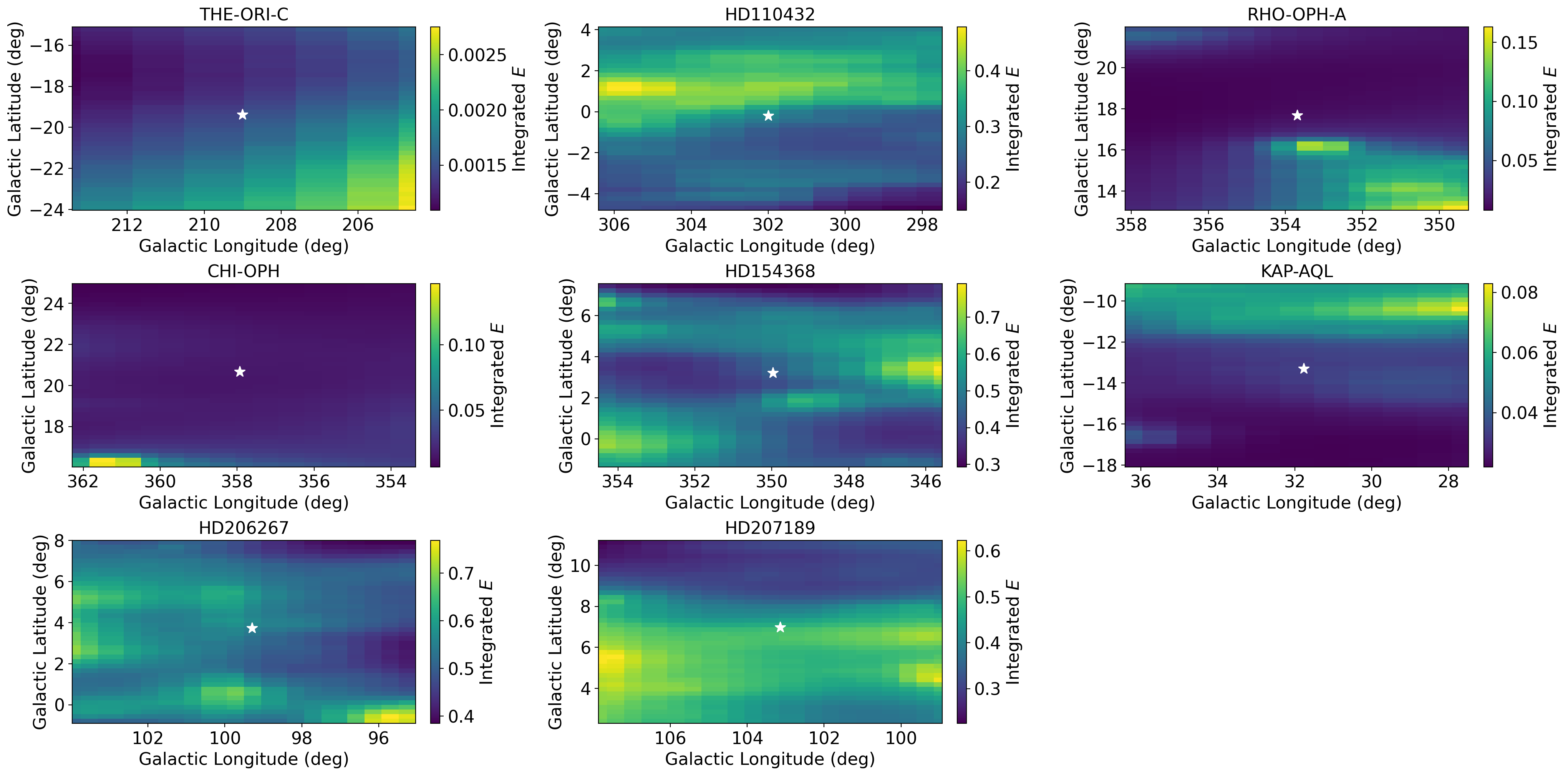}
    \caption{2D dust map in the region around each target taken from E+24.}
    \label{fig:2d-dust-maps-all-targets}
\end{figure*}

We include 2D dust maps around each of the targets in our sample in Fig.
\ref{fig:2d-dust-maps-all-targets}. These are produced by integrating the E+24
3D dust maps over the distance between the Sun and the star.  

The inner-most 69 pc of the dust extinction density curves has been removed by
E+24 because their models produce an artificial spike in dust extinction density
in the regions closest to the Sun, which is caused by incorporating the very low
extinction values from \cite{Zhang+2023} into their methodology. Since previous
work show that these nearby regions are essentially dust free
\citep{Leike+2020}, this exclusion does not heavily affect their dust maps
overall, nor the results in our paper.

This analysis is based on a single $R(\lambda)$ law, as is assumed in
\cite{Zhang+2023}. $R(V)$ represents the size of dust grains in the ISM, with
higher $R(V)$ values indicating larger dust grains. Variations around the
canonical Milky Way value of $R(\lambda) = 3.1$ across different line of sight
have been shown to be small, $\sigma(R(\lambda)) = 0.18$, with less than one
percent having $R(\lambda) > 4$ \citep{Schlafly+2016}. Different values of
$R(\lambda)$ would change $E$ by a small factor, which would not affect the
matching of absorption components and dust extinction density peaks. There could
be an impact, however, on the overall relation between dust depletion and dust
extinction density presented in Figure \ref{fig:depl-vs-ext}, where $E$ density
for different lines of sight could shift vertically by a small factor. This
could especially be the case for the gas clouds in or near the Orion Nebula,
which is known to have high dust content, i.e. those clouds along the line of
sight towards $\theta^1$ Ori C.

$\rho$ Oph A is the most nearby target. R-H+25 are not able to perform a
component-by-component analysis for this line of sight due to saturation of most
of the metal absorption lines, and are only able to constrain \znfei\, for one
of the components. We deduce that this results from there being large amounts of
gas and metal content along this line of sight. Indeed, this line of sight has a
high overall dust extinction $A_V = 2.58 \pm 0.34$ \citep{Valencic+2004}. The
dust extinction density map towards $\rho$ Oph A also has a relatively large
peak at a distance of 121.2 pc from the Sun. This shows that, in addition to
there being a lot of gas along this line of sight, there is also a significant
amount of dust.

$\chi$ Oph is the second most nearby target in our sample (after $\rho$ Oph A),
at a distance of 152 $\pm$ 5 pc. Along this line of sight is the component with
the highest amount of dust depletion ([Zn/Fe]$_{\mathrm{fit},2} = 2.06 \pm
0.16$), which we associate with the highest peak in dust extinction density at a
distance of 150.1 pc (see Fig. \ref{fig:CHI-OPH}). This peak also has one of the
largest values of dust extinction density, making this line of sight remarkably
dust and gas rich.

The line of sight towards HD~154368 has many absorption components, which is
also a feature of the dust extinction density curve (see Fig.
\ref{fig:HD154368}). The dust cloud that contributes the majority to the dust
extinction is at 201.2 pc. This is one of the densest in our sample, and we
associate it with the absorption component with the highest level of dust
depletion along this line of sight, \znfe$_{, 4} = 1.85 \pm 0.09$. 

The lines of sight towards both HD~206267 and HD~207190 have significantly more
peaks in extinction density than there are groups of absorption components.
These two lines of sight are shown to be chemically similar in R-H+25. Both
lines of sight are in regions of high dust extinction, evident from the E+24 2D
dust maps and shown in the relevant panels of Fig.
\ref{fig:2d-dust-maps-all-targets}. This means that the dust extinction density
maps likely encompass dust clouds adjacent to these lines of sight in addition
to those observed only in absorption. It is also possible that the grouping of
absorption components are overly conservative and therefore miss some of the
substructure along the lines of sight. The Ca\,\textsc{ii} absorption line of
HD~206267 is indeed broad and could have additional blended absorption lines
(see the top panel of Fig. \ref{fig:HD206267}).

\begin{figure}
    \centering
    \includegraphics[width=0.9\linewidth]{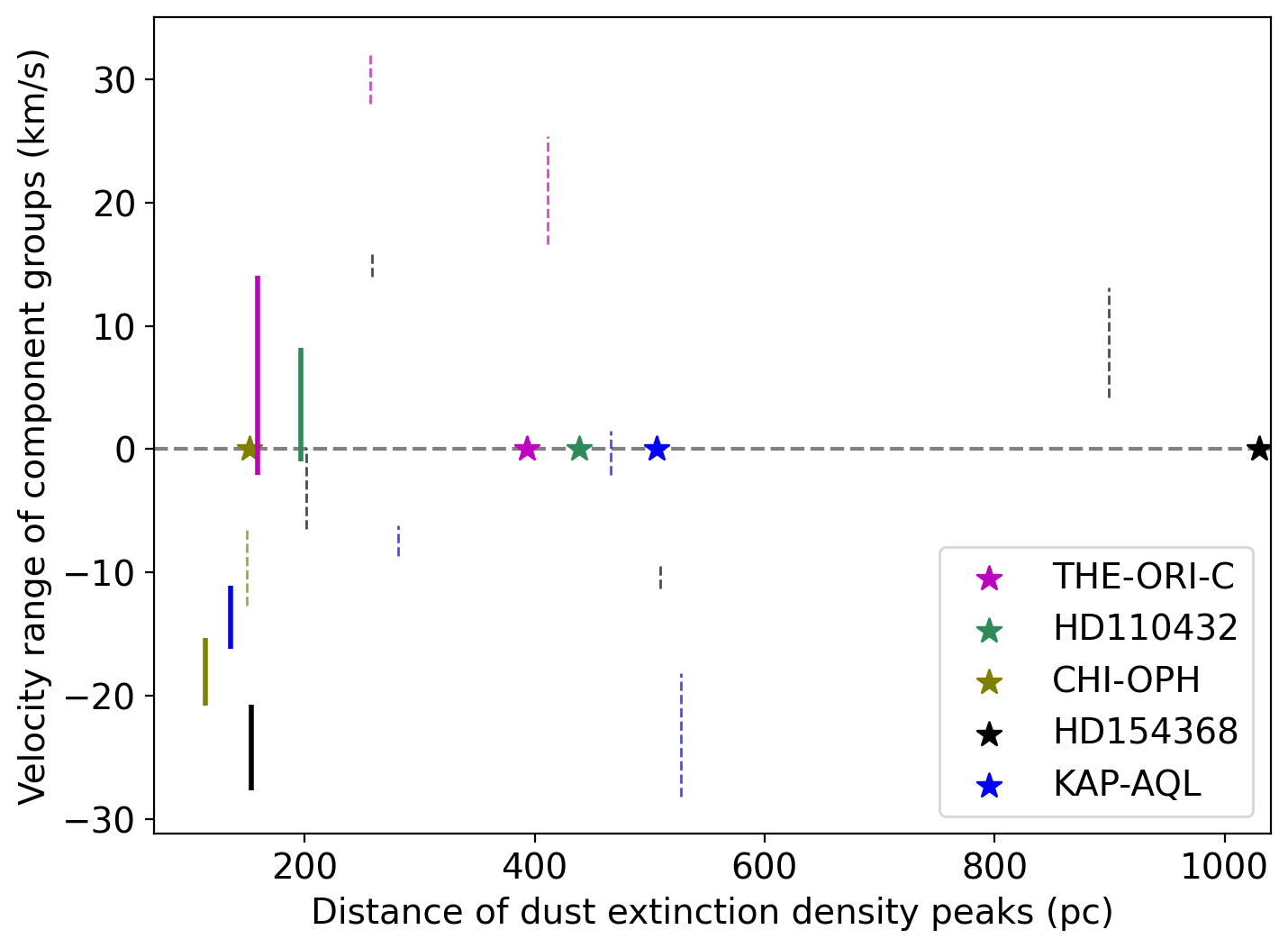}
    \caption{Velocity as a function of distance from the Sun for gas clouds along all lines of sight. The solid lines show the clouds associated with the Local Bubble, per the definition in \cite{ONeill+2024}, and dashed lines show all other clouds. The stars show their distances from the Sun.}
    \label{fig:velocity-vs-distance}
\end{figure}

3D dust extinction maps have been used to study the Local Bubble in detail,
enabling results on its geometry and origin. In particular, \cite{ONeill+2024}
use the E+24 dust map to show that the Local Bubble resembles a chimney, likely
having formed due to the bursting of the supernova-driven bubble.
\cite{ONeill+2024} define the surface of the Local Bubble as the first
significant peak in the E+24 dust extinction density maps along lines of sight.
Here, we use the results of our matching exercise to show the velocities of
these peaks as a function of distance (solid lines in Figure
\ref{fig:velocity-vs-distance}). We exclude the lines of sight towards HD~206267
and HD~207198 from this analysis because the mismatch between the number of
absorption components and dust extinction density peaks makes them difficult to
interpret. 

In Fig. \ref{fig:velocity-vs-distance} we see that the closest components along
lines of sight toward $\theta^1$ Ori C and HD~110432 exhibit negative
velocities, while more distant components toward $\rho$ Oph A, $\chi$ Oph, and
$\kappa$ Aql show less negative or positive velocities. This pattern is
consistent with a velocity gradient and can be interpreted as the Local Bubble
expanding relative to the Sun. In this frame, the Sun’s motion relative to the
local standard of rest \citep[LSR, $U_{\odot}, V_{\odot}, W_{\odot} = (11.1,
12.2, 7.2)$ km s$^{-1}$;][]{Schoenrich+2010} is toward $\theta^1$ Ori C ($-17.2$
km s$^{-1}$) and HD~110432 ($-4.5$ km s$^{-1}$), and away from $\chi$ Oph (12.5
km s$^{-1}$), HD~154368 (9.2 km s$^{-1}$) and $\kappa$ Aql (13.8 km s$^{-1}$).
Therefore, the observed negative velocities of the nearest components are partly
due to the Sun moving toward these stars, while the further components appear
less negative or positive because the motion of the Sun is less aligned along
those directions. 

\begin{figure}
    \centering
    \includegraphics[width=\linewidth]{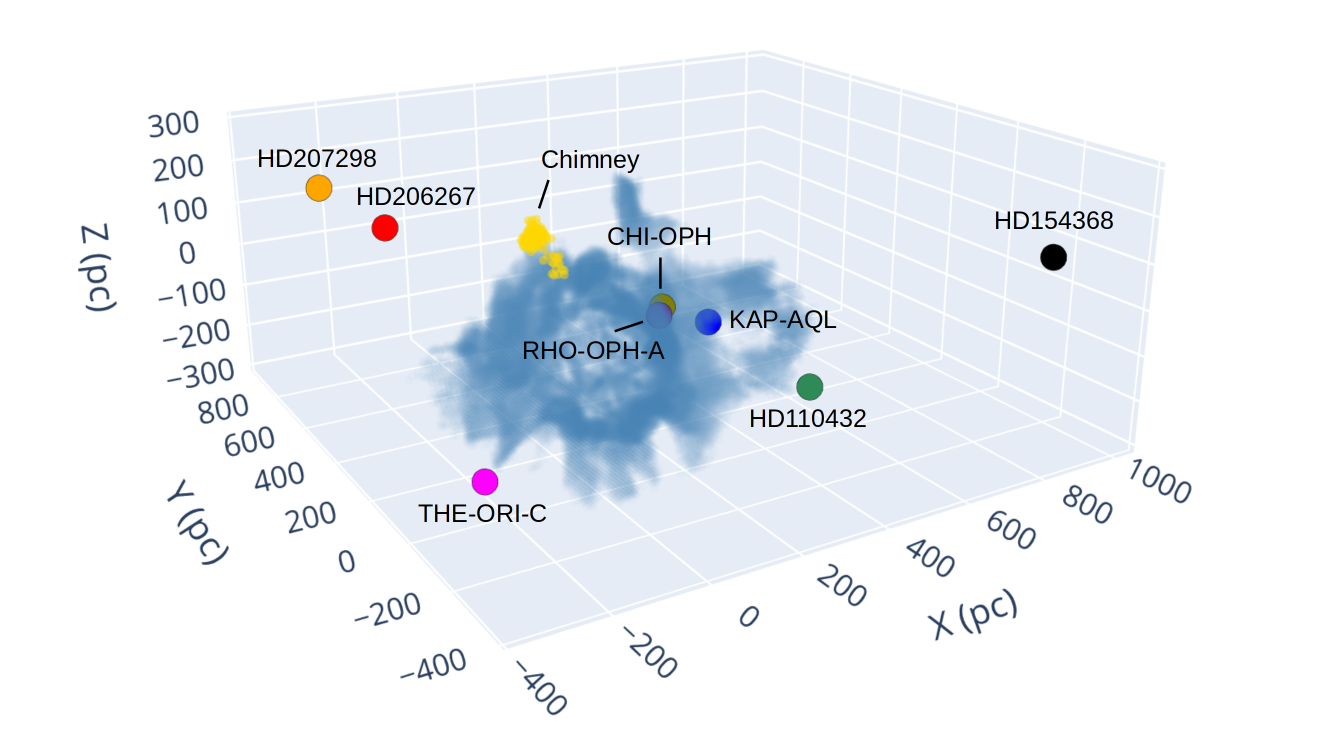}
    \caption{Three-dimensional plot of the surface of the Local Bubble as taken from \cite{ONeill+2024}, shown in blue. Our stellar targets are plotted as points and labelled. The location of the chimney structure reported in \cite{ONeill+2024} is plotted in yellow and labelled. The associated interactive 3D plot is available online.}
    \label{fig:localbubble}
\end{figure}

These results further support our methodology, showing that the way in which we
have matched absorption components to dust extinction density peaks reproduces
physical structures revealed in 3D dust maps. We include Fig.
\ref{fig:localbubble} to show the stellar targets with respect to the Local
Bubble, as it it reported in \cite{ONeill+2024}. The associated interactive 3D
plot is available online\footnote{Available at \url{https://github.com/tanita-rh/connecting-the-dusty-dots}}.

\section{Conclusions} \label{sec:conclusions}

In this paper we consider dust depletion and dust extinction as two
complementary probes of interstellar dust content in individual (groups of) ISM
clouds. We use dust depletion measurements of individual (groups of) gas clouds
along eight lines of sight within 1.1 kpc of the Sun from
\cite{Ramburuth-Hurt+2025}, and 3D dust extinction density maps from
\cite{Edenhofer+2024}. Supported by the well-known relationship between gas and
dust in the ISM \citep{Bohlin+1978, Cardelli+1989}, we assume a correlation
between dust depletion and dust extinction density. We use this correlation to
associate the absorbing clouds with peaks in dust extinction density and
pinpoint the likely locations of absorption components along each line of sight.
We show that different lines of sight lie on a similar correlation between dust
extinction density and dust depletion (see Fig. \ref{fig:depl-vs-ext}), which
suggests that this relation may be physical. We find that the location of the
highest dust extinction density peak coincides with an independent distance
measurement of the absorption component with the highest level of dust depletion
for the line of sight towards $\theta^1$ Ori C, which further corroborates our
methodology. 

We find that the number of (groups of) absorption components and the number of
peaks in the dust extinction density peaks generally agree within $\pm$ 1 -- 2
for six of eight lines of sight ($\theta^1$ Ori C, HD~110432, $\rho$ Oph A,
$\chi$ Oph, HD~154368, and $\kappa$ Aql). For five of these (excluding $\rho$
Oph A), we obtain distance estimates for absorbing gas clouds along their lines
of sight, including those associated with the Local Bubble. 

For the remaining two lines of sight (HD~206267 and HD~207198), there are many
more peaks in dust extinction density than groups of absorption components,
which is probably due to a blend of absorbing components that cannot be easily
separated and/or that the larger angular size of the dust extinction density
maps capture more dust content than the pencil-beam of absorption-line
spectroscopy.

Along the line of sight to $\theta^1$ Ori C, Components 1 and 3 differ
significantly in dust depletion by $0.56 \pm 0.25$ over a distance of 98 pc. We
interpret this as an indication of the physical scale over which chemical mixing
remains incomplete in the ISM of the Milky Way.

Our analysis of dust extinction and velocity structure supports the
interpretation of the Local Bubble as an expanding cavity surrounding the Sun.
We find that the shell of the Bubble nearest to the Sun show negative
velocities, while the further side display less negative or positive velocities.
This is consistent with an expansion-driven velocity gradient, while taking into
account the motion of the Sun relative to the local standard of rest. These
results further support our methodology of matching the absorption components
with the highest level of dust depletion to the strongest peaks in dust
extinction density along the lines of sight.

In this work, we have shown that detailed component-by-component analyses,
combining the power of absorption-line spectroscopy with 3D dust maps, provide
invaluable insights into the nature of dust in the local ISM of the Milky Way.

\begin{acknowledgements}
   T.R.-H., A.D.C., J.-K.K. acknowledge support by the Swiss National Science Foundation under grant 185692. This work is based on observations with the NASA/ESA Hubble Space Telescope obtained from Mikulski Archive for Space Telescopes at the Space Telescope Science Institute, which is operated by the Association of Universities for Research in Astronomy, Incorporated, under NASA contract NAS5-26555. This research has made use of NASA’s Astrophysics Data System.
\end{acknowledgements}

\bibliographystyle{aa}
\bibliography{AandA_MilkyWay}

@ARTICLE{Ramburuth-Hurt+2023,
       author = {{Ramburuth-Hurt}, T. and {De Cia}, A. and {Krogager}, J. -K. and {Ledoux}, C. and {Petitjean}, P. and {P{\'e}roux}, C. and {Dessauges-Zavadsky}, M. and {Fynbo}, J. and {Wendt}, M. and {Bouch{\'e}}, N.~F. and {Konstantopoulou}, C. and {Jermann}, I.},
        title = "{Chemical diversity of gas in distant galaxies. Metal and dust enrichment and variations within absorbing galaxies}",
      journal = {Astronomy \& Astrophysics},
         year = 2023,
        month = apr,
       volume = {672},
          eid = {A68},
        pages = {A68}
}

@ARTICLE{Savage&Sembach1996,
       author = {{Savage}, Blair D. and {Sembach}, Kenneth R.},
        title = "{Interstellar Abundances from Absorption-Line Observations with the Hubble Space Telescope}",
      journal = {Annual Review of Astronomy and Astrophysics},
         year = 1996,
        month = jan,
       volume = {34},
        pages = {279-330}
        }

@ARTICLE{Jenkins2009,
       author = {{Jenkins}, Edward B.},
        title = "{A Unified Representation of Gas-Phase Element Depletions in the Interstellar Medium}",
      journal = {The Astrophysical Journal},
         year = 2009,
        month = aug,
       volume = {700},
       number = {2},
        pages = {1299-1348}
}

@ARTICLE{DeCia+2016,
       author = {{De Cia}, A. and {Ledoux}, C. and {Mattsson}, L. and {Petitjean}, P. and {Srianand}, R. and {Gavignaud}, I. and {Jenkins}, E.~B.},
        title = "{Dust-depletion sequences in damped Lyman-{\ensuremath{\alpha}} absorbers. A unified picture from low-metallicity systems to the Galaxy}",
      journal = {\aap},
         year = 2016,
        month = dec,
       volume = {596},
          eid = {A97},
        pages = {A97}
}

@ARTICLE{Konstantopoulou+2022,
       author = {{Konstantopoulou}, Christina and {De Cia}, Annalisa and {Krogager}, Jens-Kristian and {Ledoux}, C{\'e}dric and {Noterdaeme}, Pasquier and {Fynbo}, Johan P.~U. and {Heintz}, Kasper E. and {Watson}, Darach and {Andersen}, Anja C. and {Ramburuth-Hurt}, Tanita and {Jermann}, Iris},
        title = "{Dust depletion of metals from local to distant galaxies. I. Peculiar nucleosynthesis effects and grain growth in the ISM}",
      journal = {\aap},
         year = 2022,
        month = oct,
       volume = {666},
          eid = {A12},
        pages = {A12}
}

@misc{Krogager2018_VoigtFit,
  author = {{Krogager}, Jens-Kristian},
  title = "{VoigtFit: Absorption line fitting for Voigt profiles}",
  howpublished = {Astrophysics Source Code Library, record ascl:1811.016},
  year = 2018,
  month = nov,
  eid = {ascl:1811.016},
  adsurl = {https://ui.adsabs.harvard.edu/abs/2018ascl.soft11016K}
}

@ARTICLE{DeCia+2021,
       author = {{De Cia}, Annalisa and {Jenkins}, Edward B. and {Fox}, Andrew J. and {Ledoux}, C{\'e}dric and {Ramburuth-Hurt}, Tanita and {Konstantopoulou}, Christina and {Petitjean}, Patrick and {Krogager}, Jens-Kristian},
        title = "{Large metallicity variations in the Galactic interstellar medium}",
      journal = {\nat},
         year = 2021,
        month = sep,
       volume = {597},
       number = {7875},
        pages = {206-208},
          doi = {10.1038/s41586-021-03780-0}
}

@ARTICLE{Ritchey+2023,
       author = {{Ritchey}, Adam M. and {Jenkins}, Edward B. and {Shull}, J. Michael and {Savage}, Blair D. and {Federman}, S.~R. and {Lambert}, David L.},
        title = "{The Distribution of Metallicities in the Local Galactic Interstellar Medium}",
      journal = {\apj},
         year = 2023,
        month = jul,
       volume = {952},
       number = {1},
          eid = {57},
        pages = {57}
        }

@ARTICLE{Green+2019,
       author = {{Green}, Gregory M. and {Schlafly}, Edward and {Zucker}, Catherine and {Speagle}, Joshua S. and {Finkbeiner}, Douglas},
        title = "{A 3D Dust Map Based on Gaia, Pan-STARRS 1, and 2MASS}",
      journal = {\apj},
         year = 2019,
        month = dec,
       volume = {887},
       number = {1},
          eid = {93},
        pages = {93},
          doi = {10.3847/1538-4357/ab5362}
          }

@ARTICLE{Lallement+2018,
       author = {{Lallement}, R. and {Capitanio}, L. and {Ruiz-Dern}, L. and {Danielski}, C. and {Babusiaux}, C. and {Vergely}, L. and {Elyajouri}, M. and {Arenou}, F. and {Leclerc}, N.},
        title = "{Three-dimensional maps of interstellar dust in the Local Arm: using Gaia, 2MASS, and APOGEE-DR14}",
      journal = {\aap},
         year = 2018,
        month = aug,
       volume = {616},
          eid = {A132},
        pages = {A132},
          doi = {10.1051/0004-6361/201832832}
          }

@ARTICLE{Gordon+2003,
       author = {{Gordon}, Karl D. and {Clayton}, Geoffrey C. and {Misselt}, K.~A. and {Landolt}, Arlo U. and {Wolff}, Michael J.},
        title = "{A Quantitative Comparison of the Small Magellanic Cloud, Large Magellanic Cloud, and Milky Way Ultraviolet to Near-Infrared Extinction Curves}",
      journal = {\apj},
         year = 2003,
        month = sep,
       volume = {594},
       number = {1},
        pages = {279-293},
          doi = {10.1086/376774},
archivePrefix = {arXiv},
       eprint = {astro-ph/0305257},
 primaryClass = {astro-ph},
       adsurl = {https://ui.adsabs.harvard.edu/abs/2003ApJ...594..279G}
}

@ARTICLE{Wiseman+2017,
       author = {{Wiseman}, P. and {Schady}, P. and {Bolmer}, J. and {Kr{\"u}hler}, T. and {Yates}, R.~M. and {Greiner}, J. and {Fynbo}, J.~P.~U.},
        title = "{Evolution of the dust-to-metals ratio in high-redshift galaxies probed by GRB-DLAs}",
      journal = {\aap},
         year = 2017,
        month = mar,
       volume = {599},
          eid = {A24},
        pages = {A24},
          doi = {10.1051/0004-6361/201629228},
archivePrefix = {arXiv},
       eprint = {1607.00288},
 primaryClass = {astro-ph.GA},
       adsurl = {https://ui.adsabs.harvard.edu/abs/2017A&A...599A..24W}
}

@ARTICLE{Rezaei+2024,
       author = {{Rezaei Kh.}, Sara and {Beuther}, Henrik and {Benjamin}, Robert A. and {Eilers}, Anna-Christina and {Henning}, Thomas and {Jim{\'e}nez-Donaire}, Maria J. and {Miville-Desch{\^e}nes}, Marc-Antoine},
        title = "{3D structure of the Milky Way out to 10 kpc from the Sun: Catalogue of large molecular clouds in the Galactic Plane}",
      journal = {\aap},
         year = 2024,
        month = dec,
       volume = {692},
          eid = {A255},
        pages = {A255},
          doi = {10.1051/0004-6361/202449255},
archivePrefix = {arXiv},
       eprint = {2405.09634},
 primaryClass = {astro-ph.GA},
       adsurl = {https://ui.adsabs.harvard.edu/abs/2024A&A...692A.255R}
}

@ARTICLE{Dharmawardena+2024,
       author = {{Dharmawardena}, T.~E. and {Bailer-Jones}, C.~A.~L. and {Fouesneau}, M. and {Foreman-Mackey}, D. and {Coronica}, P. and {Colnaghi}, T. and {M{\"u}ller}, T. and {Wilson}, A.~G.},
        title = "{All-sky three-dimensional dust density and extinction Maps of the Milky Way out to 2.8 kpc}",
      journal = {\mnras},
         year = 2024,
        month = aug,
       volume = {532},
       number = {3},
        pages = {3480-3498},
          doi = {10.1093/mnras/stae1474},
archivePrefix = {arXiv},
       eprint = {2406.06740},
 primaryClass = {astro-ph.GA},
       adsurl = {https://ui.adsabs.harvard.edu/abs/2024MNRAS.532.3480D}
}

@ARTICLE{Konstantopoulou+2024,
       author = {{Konstantopoulou}, Christina and {De Cia}, Annalisa and {Ledoux}, C{\'e}dric and {Krogager}, Jens-Kristian and {Mattsson}, Lars and {Watson}, Darach and {Heintz}, Kasper E. and {P{\'e}roux}, C{\'e}line and {Noterdaeme}, Pasquier and {Andersen}, Anja C. and {Fynbo}, Johan P.~U. and {Jermann}, Iris and {Ramburuth-Hurt}, Tanita},
        title = "{Dust depletion of metals from local to distant galaxies. II. Cosmic dust-to-metal ratio and dust composition}",
      journal = {\aap},
         year = 2024,
        month = jan,
       volume = {681},
          eid = {A64},
        pages = {A64},
          doi = {10.1051/0004-6361/202347171},
archivePrefix = {arXiv},
       eprint = {2310.07709},
 primaryClass = {astro-ph.GA},
       adsurl = {https://ui.adsabs.harvard.edu/abs/2024A&A...681A..64K}
}

@ARTICLE{Zucker+2022,
       author = {{Zucker}, Catherine and {Goodman}, Alyssa A. and {Alves}, Jo{\~a}o and {Bialy}, Shmuel and {Foley}, Michael and {Speagle}, Joshua S. and {Gro{\^I}{\texttwosuperior}schedl}, Josefa and {Finkbeiner}, Douglas P. and {Burkert}, Andreas and {Khimey}, Diana and {Swiggum}, Cameren},
        title = "{Star formation near the Sun is driven by expansion of the Local Bubble}",
      journal = {\nat},
         year = 2022,
        month = jan,
       volume = {601},
       number = {7893},
        pages = {334-337},
          doi = {10.1038/s41586-021-04286-5},
archivePrefix = {arXiv},
       eprint = {2201.05124},
 primaryClass = {astro-ph.GA},
       adsurl = {https://ui.adsabs.harvard.edu/abs/2022Natur.601..334Z}
}

@ARTICLE{Field1974,
       author = {{Field}, G.~B.},
        title = "{Interstellar abundances: gas and dust.}",
      journal = {\apj},
         year = 1974,
        month = feb,
       volume = {187},
        pages = {453-459},
          doi = {10.1086/152654},
       adsurl = {https://ui.adsabs.harvard.edu/abs/1974ApJ...187..453F}
}

@ARTICLE{Roman-Duval+2021,
       author = {{Roman-Duval}, Julia and {Jenkins}, Edward B. and {Tchernyshyov}, Kirill and {Williams}, Benjamin and {Clark}, Christopher J.~R. and {Gordon}, Karl D. and {Meixner}, Margaret and {Hagen}, Lea and {Peek}, Joshua and {Sandstrom}, Karin and {Werk}, Jessica and {Yanchulova Merica-Jones}, Petia},
        title = "{METAL: The Metal Evolution, Transport, and Abundance in the Large Magellanic Cloud Hubble Program. II. Variations of Interstellar Depletions and Dust-to-gas Ratio within the LMC}",
      journal = {\apj},
         year = 2021,
        month = apr,
       volume = {910},
       number = {2},
          eid = {95},
        pages = {95},
          doi = {10.3847/1538-4357/abdeb6},
archivePrefix = {arXiv},
       eprint = {2101.09399},
 primaryClass = {astro-ph.GA},
       adsurl = {https://ui.adsabs.harvard.edu/abs/2021ApJ...910...95R}
}

@ARTICLE{Welty+2020,
       author = {{Welty}, Daniel E. and {Sonnentrucker}, Paule and {Snow}, Theodore P. and {York}, Donald G.},
        title = "{HD 62542: Probing the Bare, Dense Core of a Translucent Interstellar Cloud}",
      journal = {\apj},
         year = 2020,
        month = jul,
       volume = {897},
       number = {1},
          eid = {36},
        pages = {36},
          doi = {10.3847/1538-4357/ab8f8e},
archivePrefix = {arXiv},
       eprint = {2005.10846},
 primaryClass = {astro-ph.GA},
       adsurl = {https://ui.adsabs.harvard.edu/abs/2020ApJ...897...36W}
}

@ARTICLE{Price+2001,
       author = {{Price}, R.~J. and {Crawford}, I.~A. and {Barlow}, M.~J. and {Howarth}, I.~D.},
        title = "{An ultra-high-resolution study of the interstellar medium towards Orion}",
      journal = {\mnras},
         year = 2001,
        month = dec,
       volume = {328},
       number = {2},
        pages = {555-582},
          doi = {10.1046/j.1365-8711.2001.04893.x},
       adsurl = {https://ui.adsabs.harvard.edu/abs/2001MNRAS.328..555P}
}

@ARTICLE{Savaglio&Fall2004,
       author = {{Savaglio}, Sandra and {Fall}, S. Michael},
        title = "{Dust Depletion and Extinction in a Gamma-Ray Burst Afterglow}",
      journal = {\apj},
         year = 2004,
        month = oct,
       volume = {614},
       number = {1},
        pages = {293-300},
          doi = {10.1086/423447},
archivePrefix = {arXiv},
       eprint = {astro-ph/0406430},
 primaryClass = {astro-ph},
       adsurl = {https://ui.adsabs.harvard.edu/abs/2004ApJ...614..293S}
}

@ARTICLE{Bolmer+2019,
       author = {{Bolmer}, J. and {Ledoux}, C. and {Wiseman}, P. and {De Cia}, A. and {Selsing}, J. and {Schady}, P. and {Greiner}, J. and {Savaglio}, S. and {Burgess}, J.~M. and {D'Elia}, V. and {Fynbo}, J.~P.~U. and {Goldoni}, P. and {Hartmann}, D.~H. and {Heintz}, K.~E. and {Jakobsson}, P. and {Japelj}, J. and {Kaper}, L. and {Tanvir}, N.~R. and {Vreeswijk}, P.~M. and {Zafar}, T.},
        title = "{Evidence for diffuse molecular gas and dust in the hearts of gamma-ray burst host galaxies. Unveiling the nature of high-redshift damped Lyman-{\ensuremath{\alpha}} systems}",
      journal = {\aap},
         year = 2019,
        month = mar,
       volume = {623},
          eid = {A43},
        pages = {A43},
          doi = {10.1051/0004-6361/201834422},
archivePrefix = {arXiv},
       eprint = {1810.06403},
 primaryClass = {astro-ph.GA},
       adsurl = {https://ui.adsabs.harvard.edu/abs/2019A&A...623A..43B}
}

@ARTICLE{Pei1992,
       author = {{Pei}, Yichuan C.},
        title = "{Interstellar Dust from the Milky Way to the Magellanic Clouds}",
      journal = {\apj},
         year = 1992,
        month = aug,
       volume = {395},
        pages = {130},
          doi = {10.1086/171637},
       adsurl = {https://ui.adsabs.harvard.edu/abs/1992ApJ...395..130P}
}

@ARTICLE{Savage&Mathis1979,
       author = {{Savage}, B.~D. and {Mathis}, J.~S.},
        title = "{Observed properties of interstellar dust.}",
      journal = {\araa},
         year = 1979,
        month = jan,
       volume = {17},
        pages = {73-111},
          doi = {10.1146/annurev.aa.17.090179.000445},
       adsurl = {https://ui.adsabs.harvard.edu/abs/1979ARA&A..17...73S}
}

@ARTICLE{Ramburuth-Hurt+2025,
       author = {{Ramburuth-Hurt}, T. and {De Cia}, A. and {Krogager}, J. -K. and {Ledoux}, C. and {Jenkins}, E. and {Fox}, A.~J. and {Konstantopoulou}, C. and {Velichko}, A. and {Dalla Pola}, L.},
        title = "{Investigating chemical variations between interstellar gas clouds in the solar neighbourhood}",
      journal = {\aap},
         year = 2025,
        month = mar,
       volume = {695},
          eid = {A14},
        pages = {A14},
          doi = {10.1051/0004-6361/202451729},
archivePrefix = {arXiv},
       eprint = {2412.18986},
 primaryClass = {astro-ph.GA},
       adsurl = {https://ui.adsabs.harvard.edu/abs/2025A&A...695A..14R}
}

@ARTICLE{Clemens1985,
       author = {{Clemens}, D.~P.},
        title = "{Massachusetts-Stony Brook Galactic plane CO survey: the galactic disk rotation curve.}",
      journal = {\apj},
         year = 1985,
        month = aug,
       volume = {295},
        pages = {422-436},
          doi = {10.1086/163386},
       adsurl = {https://ui.adsabs.harvard.edu/abs/1985ApJ...295..422C}
}

@ARTICLE{Reid+2019,
       author = {{Reid}, M.~J. and {Menten}, K.~M. and {Brunthaler}, A. and {Zheng}, X.~W. and {Dame}, T.~M. and {Xu}, Y. and {Li}, J. and {Sakai}, N. and {Wu}, Y. and {Immer}, K. and {Zhang}, B. and {Sanna}, A. and {Moscadelli}, L. and {Rygl}, K.~L.~J. and {Bartkiewicz}, A. and {Hu}, B. and {Quiroga-Nu{\~n}ez}, L.~H. and {van Langevelde}, H.~J.},
        title = "{Trigonometric Parallaxes of High-mass Star-forming Regions: Our View of the Milky Way}",
      journal = {\apj},
         year = 2019,
        month = nov,
       volume = {885},
       number = {2},
          eid = {131},
        pages = {131},
          doi = {10.3847/1538-4357/ab4a11},
archivePrefix = {arXiv},
       eprint = {1910.03357},
 primaryClass = {astro-ph.GA},
       adsurl = {https://ui.adsabs.harvard.edu/abs/2019ApJ...885..131R}
}

@ARTICLE{Edenhofer+2024,
       author = {{Edenhofer}, Gordian and {Zucker}, Catherine and {Frank}, Philipp and {Saydjari}, Andrew K. and {Speagle}, Joshua S. and {Finkbeiner}, Douglas and {En{\ss}lin}, Torsten A.},
        title = "{A parsec-scale Galactic 3D dust map out to 1.25 kpc from the Sun}",
      journal = {\aap},
         year = 2024,
        month = may,
       volume = {685},
          eid = {A82},
        pages = {A82},
          doi = {10.1051/0004-6361/202347628},
archivePrefix = {arXiv},
       eprint = {2308.01295},
 primaryClass = {astro-ph.GA},
       adsurl = {https://ui.adsabs.harvard.edu/abs/2024A&A...685A..82E}
}

@ARTICLE{GaiaCollaboration2023,
       author = {{Gaia Collaboration} and {Vallenari}, A. and {Brown}, A.~G.~A. and {Prusti}, T. and {de Bruijne}, J.~H.~J. and {Arenou}, F. and {Babusiaux}, C. and {Biermann}, M. and {Creevey}, O.~L. and {Ducourant}, C. and {Evans}, D.~W. and {Eyer}, L. and {Guerra}, R. and {Hutton}, A. and {Jordi}, C. and {Klioner}, S.~A. and {Lammers}, U.~L. and {Lindegren}, L. and {Luri}, X. and {Mignard}, F. and {Panem}, C. and {Pourbaix}, D. and {Randich}, S. and {Sartoretti}, P. and {Soubiran}, C. and {Tanga}, P. and {Walton}, N.~A. and {Bailer-Jones}, C.~A.~L. and {Bastian}, U. and {Drimmel}, R. and {Jansen}, F. and {Katz}, D. and {Lattanzi}, M.~G. and {van Leeuwen}, F. and {Bakker}, J. and {Cacciari}, C. and {Casta{\~n}eda}, J. and {De Angeli}, F. and {Fabricius}, C. and {Fouesneau}, M. and {Fr{\'e}mat}, Y. and {Galluccio}, L. and {Guerrier}, A. and {Heiter}, U. and {Masana}, E. and {Messineo}, R. and {Mowlavi}, N. and {Nicolas}, C. and {Nienartowicz}, K. and {Pailler}, F. and {Panuzzo}, P. and {Riclet}, F. and {Roux}, W. and {Seabroke}, G.~M. and {Sordo}, R. and {Th{\'e}venin}, F. and {Gracia-Abril}, G. and {Portell}, J. and {Teyssier}, D. and {Altmann}, M. and {Andrae}, R. and {Audard}, M. and {Bellas-Velidis}, I. and {Benson}, K. and {Berthier}, J. and {Blomme}, R. and {Burgess}, P.~W. and {Busonero}, D. and {Busso}, G. and {C{\'a}novas}, H. and {Carry}, B. and {Cellino}, A. and {Cheek}, N. and {Clementini}, G. and {Damerdji}, Y. and {Davidson}, M. and {de Teodoro}, P. and {Nu{\~n}ez Campos}, M. and {Delchambre}, L. and {Dell'Oro}, A. and {Esquej}, P. and {Fern{\'a}ndez-Hern{\'a}ndez}, J. and {Fraile}, E. and {Garabato}, D. and {Garc{\'\i}a-Lario}, P. and {Gosset}, E. and {Haigron}, R. and {Halbwachs}, J. -L. and {Hambly}, N.~C. and {Harrison}, D.~L. and {Hern{\'a}ndez}, J. and {Hestroffer}, D. and {Hodgkin}, S.~T. and {Holl}, B. and {Jan{\ss}en}, K. and {Jevardat de Fombelle}, G. and {Jordan}, S. and {Krone-Martins}, A. and {Lanzafame}, A.~C. and {L{\"o}ffler}, W. and {Marchal}, O. and {Marrese}, P.~M. and {Moitinho}, A. and {Muinonen}, K. and {Osborne}, P. and {Pancino}, E. and {Pauwels}, T. and {Recio-Blanco}, A. and {Reyl{\'e}}, C. and {Riello}, M. and {Rimoldini}, L. and {Roegiers}, T. and {Rybizki}, J. and {Sarro}, L.~M. and {Siopis}, C. and {Smith}, M. and {Sozzetti}, A. and {Utrilla}, E. and {van Leeuwen}, M. and {Abbas}, U. and {{\'A}brah{\'a}m}, P. and {Abreu Aramburu}, A. and {Aerts}, C. and {Aguado}, J.~J. and {Ajaj}, M. and {Aldea-Montero}, F. and {Altavilla}, G. and {{\'A}lvarez}, M.~A. and {Alves}, J. and {Anders}, F. and {Anderson}, R.~I. and {Anglada Varela}, E. and {Antoja}, T. and {Baines}, D. and {Baker}, S.~G. and {Balaguer-N{\'u}{\~n}ez}, L. and {Balbinot}, E. and {Balog}, Z. and {Barache}, C. and {Barbato}, D. and {Barros}, M. and {Barstow}, M.~A. and {Bartolom{\'e}}, S. and {Bassilana}, J. -L. and {Bauchet}, N. and {Becciani}, U. and {Bellazzini}, M. and {Berihuete}, A. and {Bernet}, M. and {Bertone}, S. and {Bianchi}, L. and {Binnenfeld}, A. and {Blanco-Cuaresma}, S. and {Blazere}, A. and {Boch}, T. and {Bombrun}, A. and {Bossini}, D. and {Bouquillon}, S. and {Bragaglia}, A. and {Bramante}, L. and {Breedt}, E. and {Bressan}, A. and {Brouillet}, N. and {Brugaletta}, E. and {Bucciarelli}, B. and {Burlacu}, A. and {Butkevich}, A.~G. and {Buzzi}, R. and {Caffau}, E. and {Cancelliere}, R. and {Cantat-Gaudin}, T. and {Carballo}, R. and {Carlucci}, T. and {Carnerero}, M.~I. and {Carrasco}, J.~M. and {Casamiquela}, L. and {Castellani}, M. and {Castro-Ginard}, A. and {Chaoul}, L. and {Charlot}, P. and {Chemin}, L. and {Chiaramida}, V. and {Chiavassa}, A. and {Chornay}, N. and {Comoretto}, G. and {Contursi}, G. and {Cooper}, W.~J. and {Cornez}, T. and {Cowell}, S. and {Crifo}, F. and {Cropper}, M. and {Crosta}, M. and {Crowley}, C. and {Dafonte}, C. and {Dapergolas}, A. and {David}, M. and {David}, P. and {de Laverny}, P. and {De Luise}, F. and {De March}, R.},
        title = "{Gaia Data Release 3. Summary of the content and survey properties}",
      journal = {\aap},
         year = 2023,
        month = jun,
       volume = {674},
          eid = {A1},
        pages = {A1},
          doi = {10.1051/0004-6361/202243940},
archivePrefix = {arXiv},
       eprint = {2208.00211},
 primaryClass = {astro-ph.GA},
       adsurl = {https://ui.adsabs.harvard.edu/abs/2023A&A...674A...1G}
}

@ARTICLE{Zhang+2023,
       author = {{Zhang}, Xiangyu and {Green}, Gregory M. and {Rix}, Hans-Walter},
        title = "{Parameters of 220 million stars from Gaia BP/RP spectra}",
      journal = {\mnras},
         year = 2023,
        month = sep,
       volume = {524},
       number = {2},
        pages = {1855-1884},
          doi = {10.1093/mnras/stad1941},
archivePrefix = {arXiv},
       eprint = {2303.03420},
 primaryClass = {astro-ph.SR},
       adsurl = {https://ui.adsabs.harvard.edu/abs/2023MNRAS.524.1855Z}
}

@ARTICLE{Green2018,
       author = {{Green}, Gregory M.},
        title = "{dustmaps: A Python interface for maps of interstellar dust}",
      journal = {The Journal of Open Source Software},
         year = 2018,
        month = jun,
       volume = {3},
       number = {26},
        pages = {695},
          doi = {10.21105/joss.00695},
       adsurl = {https://ui.adsabs.harvard.edu/abs/2018JOSS....3..695G}
}

@ARTICLE{Leike+2020,
       author = {{Leike}, R.~H. and {Glatzle}, M. and {En{\ss}lin}, T.~A.},
        title = "{Resolving nearby dust clouds}",
      journal = {\aap},
         year = 2020,
        month = jul,
       volume = {639},
          eid = {A138},
        pages = {A138},
          doi = {10.1051/0004-6361/202038169},
archivePrefix = {arXiv},
       eprint = {2004.06732},
 primaryClass = {astro-ph.GA},
       adsurl = {https://ui.adsabs.harvard.edu/abs/2020A&A...639A.138L}
}

@ARTICLE{Bailer-Jones+2021,
       author = {{Bailer-Jones}, C.~A.~L. and {Rybizki}, J. and {Fouesneau}, M. and {Demleitner}, M. and {Andrae}, R.},
        title = "{Estimating Distances from Parallaxes. V. Geometric and Photogeometric Distances to 1.47 Billion Stars in Gaia Early Data Release 3}",
      journal = {\aj},
         year = 2021,
        month = mar,
       volume = {161},
       number = {3},
          eid = {147},
        pages = {147},
          doi = {10.3847/1538-3881/abd806},
archivePrefix = {arXiv},
       eprint = {2012.05220},
 primaryClass = {astro-ph.SR},
       adsurl = {https://ui.adsabs.harvard.edu/abs/2021AJ....161..147B}
}

@ARTICLE{Cardelli+1989,
       author = {{Cardelli}, Jason A. and {Clayton}, Geoffrey C. and {Mathis}, John S.},
        title = "{The Relationship between Infrared, Optical, and Ultraviolet Extinction}",
      journal = {\apj},
         year = 1989,
        month = oct,
       volume = {345},
        pages = {245},
          doi = {10.1086/167900},
       adsurl = {https://ui.adsabs.harvard.edu/abs/1989ApJ...345..245C}
}

@ARTICLE{Bohlin+1978,
       author = {{Bohlin}, R.~C. and {Savage}, B.~D. and {Drake}, J.~F.},
        title = "{A survey of interstellar H I from Lalpha absorption measurements. II.}",
      journal = {\apj},
         year = 1978,
        month = aug,
       volume = {224},
        pages = {132-142},
          doi = {10.1086/156357},
       adsurl = {https://ui.adsabs.harvard.edu/abs/1978ApJ...224..132B}
}

@ARTICLE{Valencic+2004,
       author = {{Valencic}, Lynne A. and {Clayton}, Geoffrey C. and {Gordon}, Karl D.},
        title = "{Ultraviolet Extinction Properties in the Milky Way}",
      journal = {\apj},
         year = 2004,
        month = dec,
       volume = {616},
       number = {2},
        pages = {912-924},
          doi = {10.1086/424922},
archivePrefix = {arXiv},
       eprint = {astro-ph/0408409},
 primaryClass = {astro-ph},
       adsurl = {https://ui.adsabs.harvard.edu/abs/2004ApJ...616..912V}
}

@ARTICLE{ONeill+2024,
       author = {{O'Neill}, Theo J. and {Zucker}, Catherine and {Goodman}, Alyssa A. and {Edenhofer}, Gordian},
        title = "{The Local Bubble Is a Local Chimney: A New Model from 3D Dust Mapping}",
      journal = {\apj},
         year = 2024,
        month = oct,
       volume = {973},
       number = {2},
          eid = {136},
        pages = {136},
          doi = {10.3847/1538-4357/ad61de},
archivePrefix = {arXiv},
       eprint = {2403.04961},
 primaryClass = {astro-ph.GA},
       adsurl = {https://ui.adsabs.harvard.edu/abs/2024ApJ...973..136O}
}

@ARTICLE{Fuchs+2006,
       author = {{Fuchs}, B. and {Breitschwerdt}, D. and {de Avillez}, M.~A. and {Dettbarn}, C. and {Flynn}, C.},
        title = "{The search for the origin of the Local Bubble redivivus}",
      journal = {\mnras},
         year = 2006,
        month = dec,
       volume = {373},
       number = {3},
        pages = {993-1003},
          doi = {10.1111/j.1365-2966.2006.11044.x},
archivePrefix = {arXiv},
       eprint = {astro-ph/0609227},
 primaryClass = {astro-ph},
       adsurl = {https://ui.adsabs.harvard.edu/abs/2006MNRAS.373..993F}
}

@ARTICLE{Schoenrich+2010,
       author = {{Sch{\"o}nrich}, Ralph and {Binney}, James and {Dehnen}, Walter},
        title = "{Local kinematics and the local standard of rest}",
      journal = {\mnras},
         year = 2010,
        month = apr,
       volume = {403},
       number = {4},
        pages = {1829-1833},
          doi = {10.1111/j.1365-2966.2010.16253.x},
archivePrefix = {arXiv},
       eprint = {0912.3693},
 primaryClass = {astro-ph.GA},
       adsurl = {https://ui.adsabs.harvard.edu/abs/2010MNRAS.403.1829S}
}

@ARTICLE{Redfield&Linksy2008,
       author = {{Redfield}, Seth and {Linsky}, Jeffrey L.},
        title = "{The Structure of the Local Interstellar Medium. IV. Dynamics, Morphology, Physical Properties, and Implications of Cloud-Cloud Interactions}",
      journal = {\apj},
         year = 2008,
        month = jan,
       volume = {673},
       number = {1},
        pages = {283-314},
          doi = {10.1086/524002},
archivePrefix = {arXiv},
       eprint = {0804.1802},
 primaryClass = {astro-ph},
       adsurl = {https://ui.adsabs.harvard.edu/abs/2008ApJ...673..283R}
}

@ARTICLE{Schlafly+2016,
       author = {{Schlafly}, E.~F. and {Meisner}, A.~M. and {Stutz}, A.~M. and {Kainulainen}, J. and {Peek}, J.~E.~G. and {Tchernyshyov}, K. and {Rix}, H.-W. and {Finkbeiner}, D.~P. and {Covey}, K.~R. and {Green}, G.~M. and {Bell}, E.~F. and {Burgett}, W.~S. and {Chambers}, K.~C. and {Draper}, P.~W. and {Flewelling}, H. and {Hodapp}, K.~W. and {Kaiser}, N. and {Magnier}, E.~A. and {Martin}, N.~F. and {Metcalfe}, N. and {Wainscoat}, R.~J. and {Waters}, C.},
        title = "{The Optical-infrared Extinction Curve and Its Variation in the Milky Way}",
      journal = {\apj},
         year = 2016,
        month = apr,
       volume = {821},
       number = {2},
          eid = {78},
        pages = {78},
          doi = {10.3847/0004-637X/821/2/78},
archivePrefix = {arXiv},
       eprint = {1602.03928},
 primaryClass = {astro-ph.GA},
       adsurl = {https://ui.adsabs.harvard.edu/abs/2016ApJ...821...78S}
}

\appendix
\section{Scatter analysis} \label{appendix:scatter_analysis}
\begin{figure}
    \centering
    \includegraphics[width=0.8\linewidth]{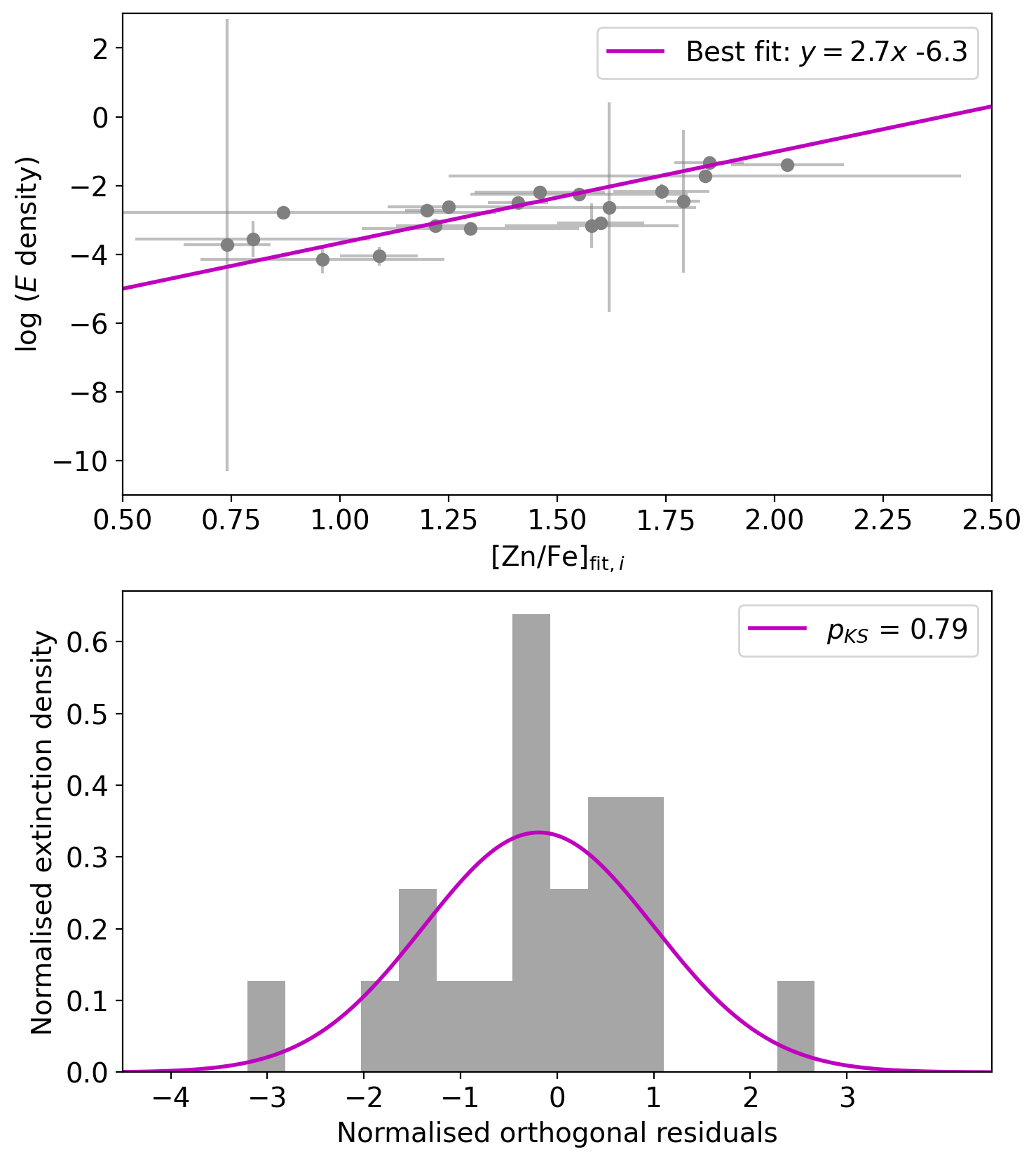}
    \caption{Top: best-fit straight line to dust extinction density vs dust depletion. The plot here is the same as Figure \ref{fig:depl-vs-ext}, with the inclusion of the best-fit line. Bottom: normalised orthogonal residuals over-plotted with its Gaussian distribution.}
    \label{fig:scatter-analysis}
\end{figure}

Since there are uncertainties on both the dust extinction density and the dust depletion, we perform an orthogonal distance regression (ODR) fit to the data and obtain a best-fit straight line of $y = mx + b$, where $m = 2.7 \pm 0.4$ and $b = -6.3 \pm 0.6$ (see the top panel of Figure \ref{fig:scatter-analysis}). We then calculate the orthogonal residuals and normalise them by the dividing by total projected uncertainties for all data points. We find that the normalised orthogonal residuals follow a Gaussian distribution, with a KS p-value of $p_{KS} = 0.79$, as shown in the bottom panel of Figure \ref{fig:scatter-analysis}. This shows that the scatter can be fully explained by the noise in the data.

\end{document}